\documentclass[a4paper,fleqn,usenatbib]{mnras}

\usepackage[T1]{fontenc}
\usepackage{ae,aecompl}

\usepackage{graphicx}	
\usepackage{amsmath}	
\usepackage{amssymb}	

\title[Long-term FRII jet evolution]{Long-term FRII jet evolution: Clues from three-dimensional simulations}

\author[Perucho, Mart\'{\i}, Quilis]{
Manel Perucho,$^{1,2}$\thanks{E-mail: manel.perucho@uv.es}
Jos\'e-Mar\'{\i}a Mart\'{\i}$^{1,2}$, Vicent Quilis$^{1,2}$
\\
$^{1}$Departament d'Astronomia i Astrof\'{\i}sica, Universitat de Val\`encia, C/ Dr. Moliner, 50, 46100, Burjassot, Val\`encia, Spain.\\
$^{2}$Observatori Astron\`omic, Universitat de Val\`encia, C/ Catedr\`atic Jos\'e Beltr\'an 2, 46980, Paterna, Val\`encia, Spain.
}

\date{Accepted XXX. Received YYY; in original form ZZZ}

\pubyear{2015}

\begin{document}
\label{firstpage}
\pagerange{\pageref{firstpage}--\pageref{lastpage}}
\maketitle

\begin{abstract}
We present a long-term numerical three-dimensional simulation of a relativistic outflow designed to be compared with previous results from axisymmetric, two-dimensional simulations, with existing analytical models and state-of-art observations. We follow the jet evolution from 1~kpc to 200~kpc, using a relativistic gas equation of state and a galactic profile for the ambient medium. We also show results from smaller scale simulations aimed to test convergence and different three-dimensional effects. We conclude that jet propagation can be faster than expected from axisymmetric simulations, covering tens of kiloparsecs in a few million years, until the dentist drill effect produced by the growth of helical instabilities slows down the propagation speed of the jet head. A comparison of key physical parameters of the jet structure as obtained from the simulations with values derived from observations of FRII sources reveals good agreement. Our simulations show that shock heating can play a significant role in the feedback from active galaxies, confirming  previous 2D results. A proper description of galactic jets as a relativistic scenario, both dynamical and thermodynamical, reveals an extremely fast and efficient feedback process reheating the ICM, and therefore, with dramatic consequences on the galactic evolution. Our results point towards FRII jets as the source of the energetic electrons observed in radio relics.
\end{abstract}




\section{Introduction}

Radio bright Active Galactic Nuclei (AGN) are currently understood as the consequence of accretion onto a supermassive black hole at the centre of the active galaxy. State-of-the-art numerical simulations that follow analytical models \citep{bz77} suggest that jets are formed when the magnetic field extracts rotational energy from the black hole \citep[e.g.,][]{tch11,tch15,po13}. Once formed in the surroundings of the central supermassive black hole (SMBH, $\sim 1-10$ gravitational radii), powerful jets propagate along several orders of magnitude in distance until the impact site with the ambient (interstellar or intergalactic) medium at distances of tens to hundreds of kiloparsecs. The jet is accelerated to relativistic velocities within the inner kiloparsecs \citep[see, e.g.][]{li09,ho15}. There exists a morphological dichotomy in AGN (radio) jet population \cite{fr74} at large scales, with a fraction of the radio source population appearing edge brightened and well collimated up to hundreds of kpc, where they show bright hot-spots (FRII type), and a majority of jets with brighter cores and decollimated, irregular structure from the inner few kpc (FRI type). FRII jets are probably mildly relativistic still at large scales, as indicated by the brightness asymmetry between jet and counter-jet \citep[e.g., Cyg A,][]{cb96}, whereas FRI jets \citep[e.g., 3C~31,][]{lb02} seem to be efficiently decelerated between 1 and 3~kpc from the central engine \citep[see, e.g.][]{bi84,la93,la96,lb14} and thus show notable symmetry at kpc-scales.      

The high degree of collimation of FRII jets can be associated either to their relativistic speeds, which implies opening angles of the order of the inverse of the flow Lorentz factor  \citep[e.g.,][]{ko12}, or to magnetic tension if the jet is surrounded by a toroidal magnetic field, or to pressure confinement by the overpressured shocked jet gas (cocoon). In addition, it has been shown that colder and faster jets develop disruptive instabilities, such as the Kelvin-Helmholtz instability, with very long growth-lengths \citep[i.e., slow growth-rates, e.g.][]{pe10}. This means that kinetically dominated (cold/fast) jets are less prone to disruption by the growth of instabilities, unlike Poynting flux dominated jets with a strong toroidal component, which are current-driven unstable \citep[e.g.,][]{mb09,mi09}.     

Numerical simulations and observations seem to indicate that jet evolution is primarily determined by the jet kinetic power \citep{rs91}, with a border at kinetic luminosities in the range $10^{44-45}\,{\rm erg/s}$. At intermediate powers, the relation between the jet and the ambient medium may be crucial for kpc-scale morphology \citep[see, e.g.][]{gw01}, as it seems to be indicated by hybrid morphology radio sources \citep[HYMORS][]{gw00}.  

A number of analytical models have been developed to track the physical parameters of powerful radio sources with time. The models are typically based on the assumption that the radio source develops through a galactic atmosphere, far from the galactic core \citep[e.g.,][]{gkw87,bc89,ka97,kf98,sch02,pm07,ka08,ka09,ma14}. \cite{ka97} (and all posterior models based on this paper) also include the assumption of self-similar expansion. The extended Begelman \& Cioffi model \citep{bc89,sch02} has proven to successfully reproduce the evolution of axisymmetric RHD, adiabatic, numerical simulations \citep{sch02,pm07,pe11}. 

Numerical simulations of relativistic jet evolution have been routinely produced since the nineties \citep{ma97,kf98}. These simulations were initially two dimensional and axisymmetric, with the jet evolving through a homogeneous ambient medium. Later, three-dimensional simulations were run \citep[e.g.][]{aloy99}, improved equations of state were added \citep{sch02,pm07}, and magnetic fields were also included in 2D simulations \citep{ko99,lei05}. Currently, 3D relativistic simulations of jet evolution, either for magnetised and non-magnetised jets are common in the literature, and with different aims, such as the study of jet evolution and structure \citep{mi10,ehk16}, the FRI-FRII dichotomy \citep[e.g.,][]{ro08,tch16}, or the feedback role of jets in their host galaxies \citep{wb11,wbu12,mu16,mu18a,mu18b,bi18,cie18}. However, the limitations imposed in these simulations by the computing demands difficult the study of long-term evolution with adequate resolution and/or grid size. In general, the aforementioned simulations have shown that numerical simulations are able to reproduce the gross large-scale structure of radio-sources, that long-wavelength instabilities can play a significant role in jet disruption \citep[although the modeling of FRI jets seems to indicate that deceleration is continuous and probably produced by small-scale mass-loading, see][]{lb14}, and that the host galaxy is certainly affected by the passage of the shock triggered by the jet as it expands. 

In previous papers \citep{pe11,pe14}, we have focused on the evolution of two-dimensional, axisymmetric jets with powers between $10^{44}\,{\rm erg/s}$ and $10^{46}\,{\rm erg/s}$ and their impact on the environment. The main conclusions of those papers are: 1) kinetic heating of the interstellar and intergalactic media by powerful extragalactic jets is   
mainly driven by the transfer of the injected energy into the ambient-medium by a strong shock, 2) ambient heating is related to the amount of energy flux carried by the jet that can be transferred, and 3) the overpressure of the shocked region can last long beyond the end of the active phase and sustain weak shocks that prevent buoyant expansion because it is the shock that drives it.  Actually, the presence of shocks around radio sources has been reported in different objects like Hercules~A \citep{nu05}, Hydra~A \citep{si09b}, MS0735.6+7421 \citep{mc05}, HCG~62 \citep{git10}, 3C~444 \citep{cro11} or PKS B1358-113 \citep{sta14}. In this respect, \citet{sha11} also claim, based on the colour evolution of galaxies in the local Universe ($z < 0.2$), that FRII kinetic feedback can be very relevant for the host galaxy evolution.
 
\cite{pe17} have shown that an appropriate description of the pressure evolution of radio lobes (cocoon in the models) and of the interaction between the jet and the ambient medium requires a relativistic description of the jet, both from a numerical and from an analytical approach. The relativistic nature of jets allows that a significant fraction of the energy flux is in the form of internal and (relativistic) kinetic energy, which can be exchanged with the ambient medium. Both this fact, and the short time scales implied by strong shocks mediating the energy exchange, make the ambient heating by relativistic jets very efficient.
 
 The jet formation at the most inner parts of galaxies is intimately connected with its effects on the interstellar and the intracluster medium, therefore, playing a crucial role shaping the host galaxies and galaxy clusters. It is nowadays commonly accepted that AGN feedback related sources must be taken into account in order to properly describe the observed properties of galaxies and clusters. The so called cooling flow problem associated to the overcooling of gas in galaxy clusters  \citep[see, e.g.,][and references therein]{qui01,mn07,fb12} can be alleviated by extracting energy from the galaxy centres to the intracluster medium (ICM). Highly linked to this scenario is the amount of cold gas feeding the galaxies from the ICM, which has direct implications on the star formation history of galaxies. Thus,  galaxy formation models neglecting the AGN heating lead to an overproduction of stars in massive galaxies \citep{oser10,lackner12, navarro13}, which results in galaxies overly massive and with a star formation  artificially extended in time. The implementation of self-consistent subgrid models of AGN  feedback  have alleviated the tension between the star formation histories and stellar contents of simulated and observed galaxies \citep[e.g.,][]{sijacki07, dub10, ga12, ga13,dub13, Vogel14,Furlog15,yang16,bourne17,Henden18}. {In this context, where the role of the AGN has become essential to understand the formation and evolution of galaxies and galaxy clusters, the correct modeling of jets, and all their associated effects, is required.} 

In this paper, we focus on the evolution of an FRII type jet with kinetic power $L_k=10^{45}\,{\rm erg/s}$ via 3D simulations. The simulated jet is equivalent to model J1 in \cite{pe11} (PQM11 hereafter) and model J45l in \citet{pe14} (PMQR14). In contrast to 2D simulations, and owing to computing constraints associated to 3D simulations, we focus on the active phase of the jet until it reaches $200\,{\rm kpc}$. Nevertheless, the simulation presented here represents the largest 3D numerical simulation devoted to jet evolution so far, in terms of grid physical size and number of cells involved. The aims of this work are to study the differences in the evolution of a jet in 2D and 3D numerical simulations, to test the results obtained in 2D regarding the efficiency of ambient heating, and to compare the evolution of the 3D jet with the evolution predicted by the extended Begelman \& Cioffi \citep[eBC,][]{sch02,pm07} analytical model. In order to do this, we used an upgraded version (see Sect.~\ref{sec:setup}) of the code \emph{Ratpenat} in the supercomputer Mare Nostrum (Barcelona Supercomputing Centre) and Tirant (Universitat de Val\`encia). The paper is structured as follows: the setup of the simulations, together with a description of the equations solved by our code are presented in Section~2. The results of the simulations are given in Section~3. Sections~4 and 5 include the discussion and the summary of this work, respectively.
 

\section{Numerical simulation} \label{sec:setup}

\subsection{The code}
\label{ss:code}

   We have used an upgraded version of our code \emph{Ratpenat} that improves the parallelization efficiency of the code thanks to a new decomposition of the numerical grid. {\it Ratpenat} is a hybrid parallel code  -- MPI + OpenMP -- that solves the  equations of relativistic hydrodynamics in conservation form using high-resolution-shock-capturing methods \citep[see][and references therein]{pe10}: i) primitive variables within numerical cells are reconstructed using PPM routines, ii) numerical fluxes across cell interfaces are computed with Marquina flux formula, iii) advance in time is performed with third order TVD-preserving Runge-Kutta methods. The upgrade that we have used permits a grid decomposition in cubes or parallelepipeds cut along all three directions (as opposed to our previous version, which only allowed us to split the grid along one of the three directions). We have used IDL software and LLNL VisIt \citep{visit} for visualisation.  
     
  The equations that are solved by the code are those corresponding to the conservation of mass, momentum and energy. These conservation equations are, for Cartesian coordinates, using units in which $c=1$: 
\begin{equation}
  \frac{\partial \mathbf{U}}{\partial t} + \frac{\partial
\mathbf{F}^x}{\partial x} + \frac{\partial \mathbf{F}^y}{\partial y} + \frac{\partial \mathbf{F}^z}{\partial z} =
\mathbf{S} ,
\end{equation}
with the vector of unknowns
\begin{equation}
  \mathbf{U}=(D,D_l,S^x,S^y,S^z,\tau)^T ,
\end{equation}
fluxes
\begin{equation}
  \mathbf{F}^i=(D v^i , D_l v^i , S^x v^i + p \delta^{xi}  , S^y v^i + p \delta^{yi}, S^z v^i + p \delta^{zi}, S^i - D v^i)^T ,
\end{equation}
with $i\, = x,\, y,\, z$, and source terms

\begin{equation}
  \mathbf{S}  =  (0, 0, g^x, g^y, g^z, v^x g^x + v^y g^y + v^z g^z)^T .
\end{equation}
 
 The six unknowns $D,D_l,S^x,S^y,S^z$ and $\tau$, refer to the densities of six conserved quantities, namely the total and leptonic rest masses, the three components of the momentum, and the energy (excluding rest-mass energy). They are all measured in the laboratory frame, and are related to the quantities in the local rest frame of the fluid (primitive variables) according to
\begin{equation}
  D = \rho W,
\end{equation}
\begin{equation}
  D_l = \rho_l W,
\end{equation}
\begin{equation}
  S^{x,y,z} = \rho h W^2 v^{x,y,z},
\end{equation}
\begin{equation}
  \tau=\rho h W^2\,-\,p\,-\,D,
\end{equation}
where $\rho$ and $\rho_l$ are the total and the leptonic rest-mass densities, respectively, $v^{x, y, z}$ are the components of the velocity of the fluid, W is the Lorentz factor ($W = 1/\sqrt{1-v^i v_i}$, where summation over index $i=x,y,z$ is implied), and $h$ is the specific enthalpy defined as
\begin{equation}
  h = 1 + \varepsilon + p/\rho,
\end{equation}
where $\varepsilon$ is the specific internal energy and $p$ is the pressure. Quantities $g^{x,y,z}$  in the definition of the source-term vector ${\bf S}$, are the components of an external gravity force that keeps the atmosphere in equilibrium (see Section~\ref{sec:su}).

The system is closed by means of the Synge equation of state \citep[][described in Appendix A of \cite{pm07}]{sy57} that accounts for a mixture of relativistic Boltzmann gases (in our case, electrons, positrons and protons). The code also integrates an equation for the jet mass fraction, $f$. This quantity, set to 1 for the injected beam material and 0 otherwise, is used as a tracer of the jet material through the grid. In these simulations, cooling has not been taken into account, as the typical  cooling times in the environment are long compared to the simulation times \citep[see Figure 10 in][]{hr02}. 

\subsection{Set up} 
\label{sec:su}

    The 3D simulation presented in this paper is initially set up as a grid filled by the ambient gas with a density and temperature profiles equivalent to those used in PQM11 and PMQR14, as derived from the modellisation of the X-ray observations of the radio galaxy 3C~31 \citep{hr02}: 
\begin{eqnarray}\label{next}
  n_{\rm ext} = n_{\rm c} \left(1 +
\left(\frac{r}{r_{\rm c}}\right)^2\right)^{-3\beta_{\rm atm,c}/2} +  \nonumber \\
+ n_{\rm g} \left(1 + \left(\frac{r}{r_{\rm g}}\right)^2\right)^{-3\beta_{\rm atm,g}/2},
\end{eqnarray}
with $r=\sqrt{x^2+y^2+z^2}$ the radial spherical coordinate, $n_{\rm c} = 0.18$~cm$^{-3}$, $r_{\rm c} = 1.2$~kpc,  $\beta_{\rm atm,c} = 0.73$, $n_{\rm g} =0.0019$~cm$^{-3}$, $r_{\rm g} = 52$~kpc, and  $\beta_{\rm atm,g} = 0.38$. The corresponding temperature profile is:

\begin{equation}\label{text}
T_{\rm ext} = \left\{ \begin{array}{ll}
T_{\rm c} + (T_{\rm g} - T_{\rm c}) \displaystyle{\frac{r}{r_{\rm m}}}, & {\rm for\,} r\leq r_{\rm m} \\
T_{\rm g}, &  {\rm for\,} r > r_{\rm m}
\end{array} \right.
\end{equation}
where $T_{\rm c}$ and $T_{\rm g}$ are characteristic temperatures of the host galaxy and the group ($4.9 \times 10^6$ K and $1.7 \times 10^7$ K, respectively), and $r_{\rm m} = 7.8$ kpc is the matching radius. This profile only accounts for the hot gas component. The cold gas component would imply an increase of the ambient density within the inner region of the host galaxy \citep[e.g.,][]{wbu12}, but is not expected to be relevant farther out, in the region actually covered by our simulations and out of the galaxy. The external pressure is derived from the number density and temperature profiles assuming a perfect gas composed of ionized hydrogen \citep{hr02,pm07}:
\begin{equation}\label{pext}
  p_{\rm ext} = \frac{k_{\rm B} T_{\rm ext}}{\mu X} n_{\rm ext},
\end{equation}
where $\mu= 0.5$ is the mass per particle in atomic mass units, $X=1$ is the abundance of hydrogen per mass unit, and $k_{\rm B}$ is the Boltzmann's constant. 

 The ambient medium is kept in equilibrium by means of a restoring force (an external gravity) as it was done in \cite{gm97,pm07,pe11} and PMQR14, which enters into the hydrodynamical equations as source terms in the momentum and energy equations. The dark matter halo accounting for the external gravity can be fitted by a NFW density profile \citep{nfw97}. All these parameters represent a moderate size galaxy cluster with mass $10^{14}\,M_{\odot}$ and $\sim 1\, \rm{Mpc}$ virial radius. In PQM11, the increase in the potential energy of the jet and ambient plasmas inflating the cavity in the gravitational well of the dark matter halo was estimated to be about one thousandth of the total energy budget brought into play by jets in standard episodes of galactic activity.

   We have chosen the jet parameters at the injection to be the same as those in simulation J45l of PMQR14, i.e., a jet with kinetic power $L_{\rm k}=10^{45}\,{\rm erg\,s^{-1}}$, injected at 1~kpc, with a radius, $R_{\rm j}=100$~pc. The flow velocity at injection is $v_{\rm j}=0.984\,c$, and the density ratio between the jet material and environment of $\rho_{\rm j}/\rho_{\rm a,0} = 5\times 10^{-4}$, resulting in $\rho_{\rm j}=8.3\times 10^{-29}$~g/cm$^3$ ($\rho_{\rm a,0} = 1.5\times10^{-25}$~g/cm$^3$), with the jet composed by electron/positron pairs at injection. In order to introduce 3D effects, we added a helical perturbation by introducing oscillatory pattern to the normal components of the jet velocity at injection. The perturbation is composed by four different frequencies spanning across two orders of magnitude, $w_1=0.01\,c/R_{\rm j}$, $w_2=0.05 \,c/{\rm R_{\rm j}}$, $w_3=0.1\,c/R_{\rm j}$, and $w_4=0.5\,c/R_{\rm j}$, with the aim to generate structures with different characteristic scales ($\lambda_i \sim 2\pi c/w_i \simeq 600,\,120,\,60,\,12\, R_{\rm j}$), and are given the same amplitude:            
 
\begin{eqnarray}
   v_{x}=v_{\rm j} \, \left( 2.5\times10^{-4}\, \sum_{i=1}^{4} \cos(w_i\,t)\right) \nonumber \\
   v_{z}=v_{\rm j} \, \left(2.5\times10^{-4}\, \sum_{i=1}^{4} \sin(w_i\,t) \right). 
\end{eqnarray}
The simulation box is split into a number of cells with a resolution of 1 cell per jet radius ($100$~pc). This is the same resolution used for the 2D simulation and it was selected on the following basis: 1) the opening angle of the jet implies a better resolution as it expands, and 2) we are mainly interested in the evolution of the global quantities related to the shocked regions. The simulation box was increased in all three directions as the bow-shock approached the current limits up to a final size of $1024\times2048\times1024$ cells in the $x$, $y$, and $z$ coordinates, respectively. This represents a physical size $\simeq 100\times200\times100$~kpc. This box size represents only a half of the distance covered by the 2D axisymmetric jet during its active phase, but it is, on the one hand, limited by the size of the 3D grid and, on the other hand, enough to compare the jet evolution during the first million years of evolution in both cases. This simulation is named J0 hereafter. We have performed several runs with increased resolution for smaller grids, for 2 (J2), 4 (J4) and 8 (J8) cells per jet radius, for grids with a physical size of $6.4\times 12.8\times 6.4$~kpc, $9.6\times 25.6\times 9.6$~kpc and $3.2\times 12.8\times 3.2$~kpc, respectively, to be used as convergence tests. We have also run a final test using the same resolution as in the main run, but decreasing the initial perturbation amplitude by a factor $10^2$ (named J0b). The physical size of this run was $25.6\times51.2\times 25.6$~kpc. 

 Using a mere cell per jet radius at injection is not standard in numerical simulations of jets. However, long-term, three-dimensional simulations as those described in this paper are exceedingly computationally demanding. The issue of the numerical convergence of our simulations and its implications on the robustness of the conclusions derived in the present paper have been addressed in Appendix~A.

    The boundary conditions are outflow at all boundaries, but at $y=0$, where injection conditions are established at the centre of this plane, at the cells through which the jet is introduced in the grid, and a reflecting boundary condition elsewhere in this plane, in order to mimic the presence of a symmetric counter-jet. The simulation was run in Mare Nostrum, at the Barcelona Supercomputing Centre, using 64 threads, adding up to a total of 1024 cores. The analysis of the results presented in the next section was run in Tirant, at the Servei d'Inform\`atica de la Universitat de Val\`encia, using 64 threads, and a total of 256 cores in this case. We used a new hybrid MPI/OMP program \emph{Mussol}, which has been developed to read the outputs generated by the new version of \emph{Ratpenat} and compute the magnitudes shown in this work.   

%
\begin{figure} 
\includegraphics[width=0.95\columnwidth]{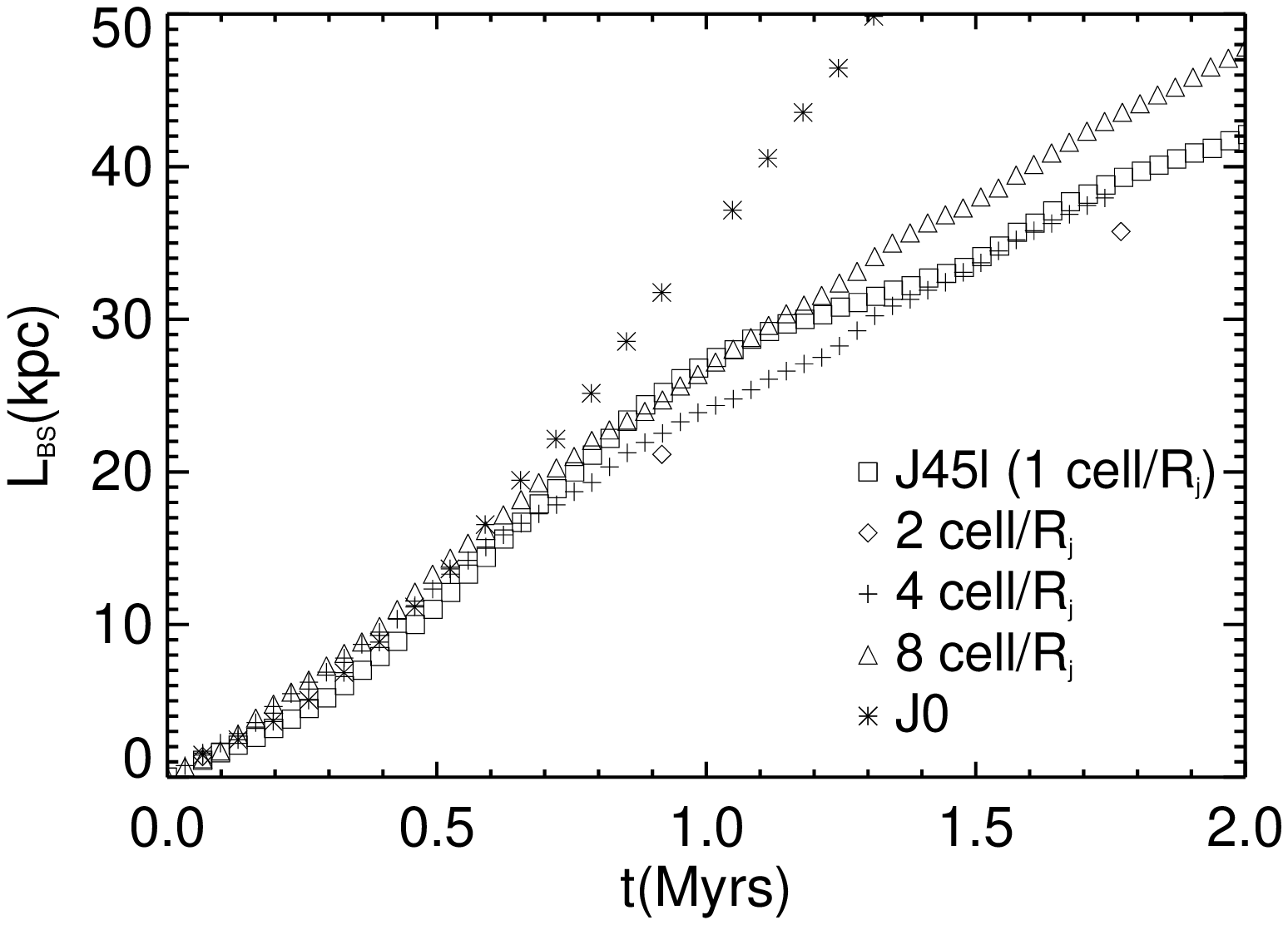}
\includegraphics[width=0.95\columnwidth]{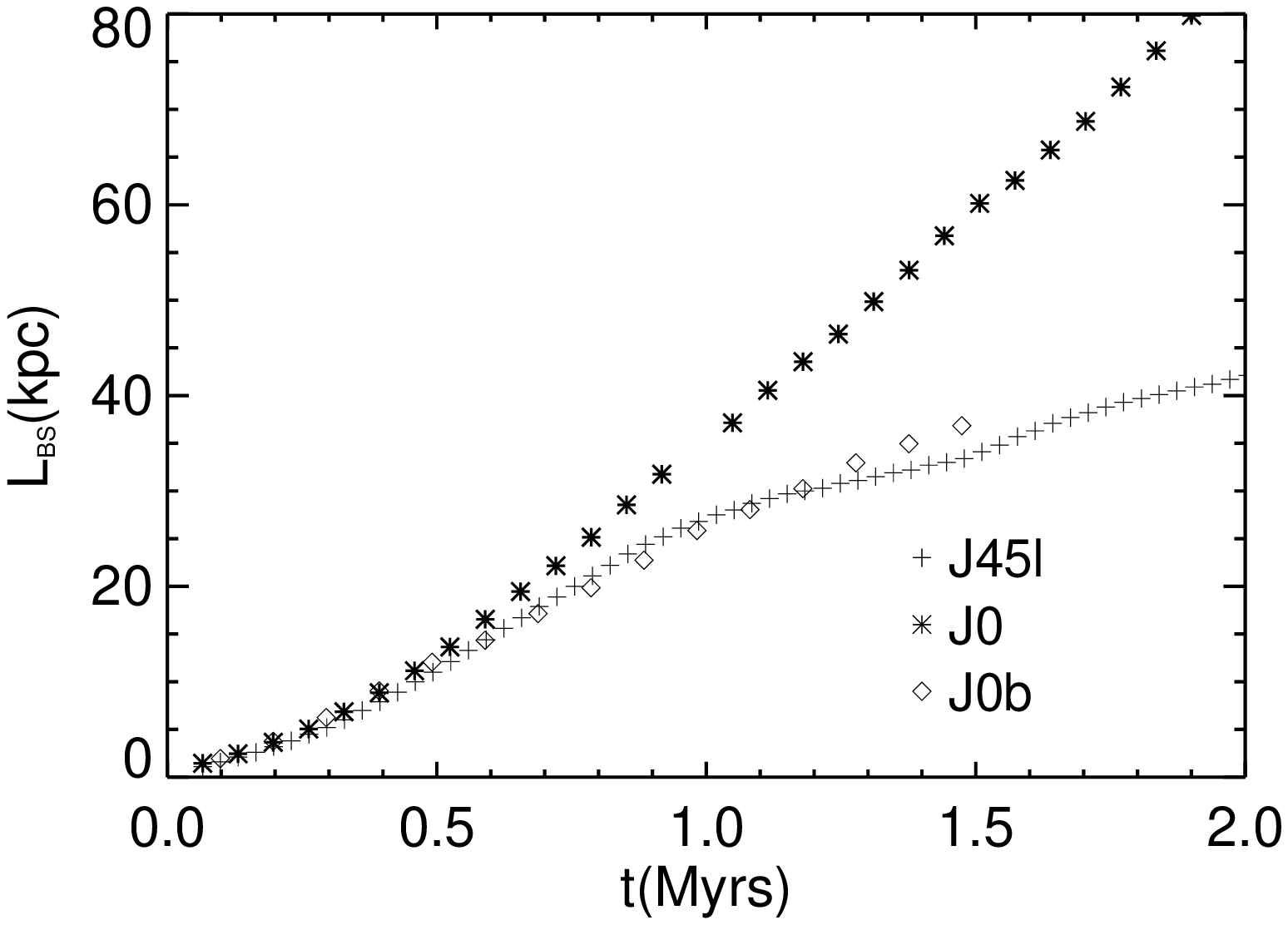}
\includegraphics[width=0.95\columnwidth]{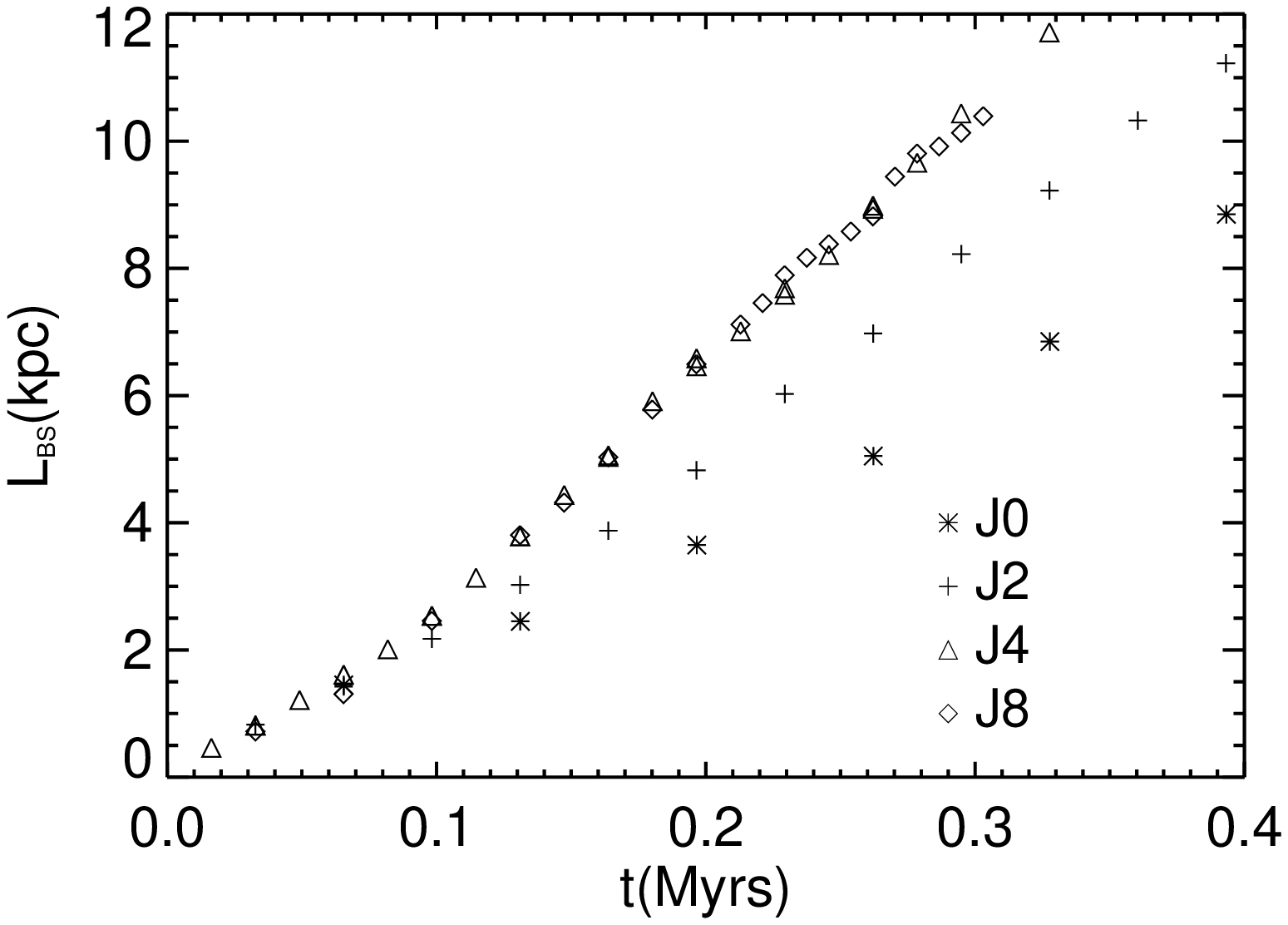}
\caption{Time evolution of the bow-shock tip position for the initial phase of the simulations presented and/or discussed in this paper. Top panel: several 2D simulations with different resolution and J0 (asterisks). Central panel: J0, J0b (smaller perturbation amplitudes), and J45l simulation from PMQR14. Bottom panel: four 3D simulations with different resolutions.}
\label{fig:inner}
\end{figure}
%

\section{Results} 
\label{sec:res}

\subsection{Previous 2D simulations}
\label{ss:2dsim}

In PQM11 we presented an axially-symmetric 2D simulation, J1 (J45l in PMQR14), of a relativistic jet with the same injection parameters and the same environment as the one discussed here. The simulation presented in PQM11 belonged to a set of simulations of jets with varying powers and compositions, aimed to investigate the characteristics of the spatial and temporal energy deposition of the jet into the ICM. Our simulations showed that heating by AGN jets is mainly shock-driven, which results in a very fast and efficient process, as compared to more gentle processes like turbulent mixing \citep[see, e.g.,][and references therein]{mn07,fb12,mn12,ga13}. In a subsequent paper (PMQR14) we found quantitative and qualitative differences of the long-term energy deposition process between non-relativistic and relativistic jets. The origin of these differences has been analyzed in \citet{pe17}. 

The long-term propagation of the jet through the ambient medium generates a characteristic and well understood morphology, with a terminal or reverse shock at the head of the jet where the flow decelerates and heats; the cocoon, a hot and light region shrouding the jet, inflated by the shocked jet particles; and a dense shell of shocked ambient medium. All these morphological elements appearing in the simulations of powerful jets have their observational counterparts. The hotspots observed at the edge of FRII jets correspond to the terminal shocks at the head of the jets. The observational counterparts of the cocoons are the extended radio lobes surrounding the jets, also associated to the X-ray cavities around many powerful radio galaxies. The leading edge of the outer shell is a bow shock which propagates through the ambient medium, also observed in some sources \citep[see, e.g.,][]{cro11,sta14}. Both cocoon and shocked ambient medium are separated by a contact discontinuity. In the following, we shall talk without distinction about {\it cocoon}, {\it cavity} or {\it lobe} when referring to the region filled with the shocked jet matter (i.e., the cocoon in the numerical simulations). Analogously, we shall talk about {\it shocked ambient medium}, {\it cavity's shell} or simply {\it shell} when referring to the region encompassing the shocked ambient matter.

The shock-dominated supersonic phase of the expansion of these cavities can be described as undertaking two consecutive stages: (i) a short {\it one-dimensional} phase governed by the one-dimensional evolution of the jet, and (ii) a genuinely {\it multidimensional} phase driven by a decelerating jet expansion as a result of the multidimensional effects affecting the jet propagation. This second phase may be followed by another one once the jets are switched-off, in which the cavities expand passively. In PQM11 the evolution of the simulations (including J1) through the one-dimensional, multidimensional and passive phases was studied and contrasted against a simple analytical
model (see also Section~\ref{ss:am}). 

\subsection{Early evolution}

%
\begin{figure*}  
\includegraphics[width=1.\columnwidth]{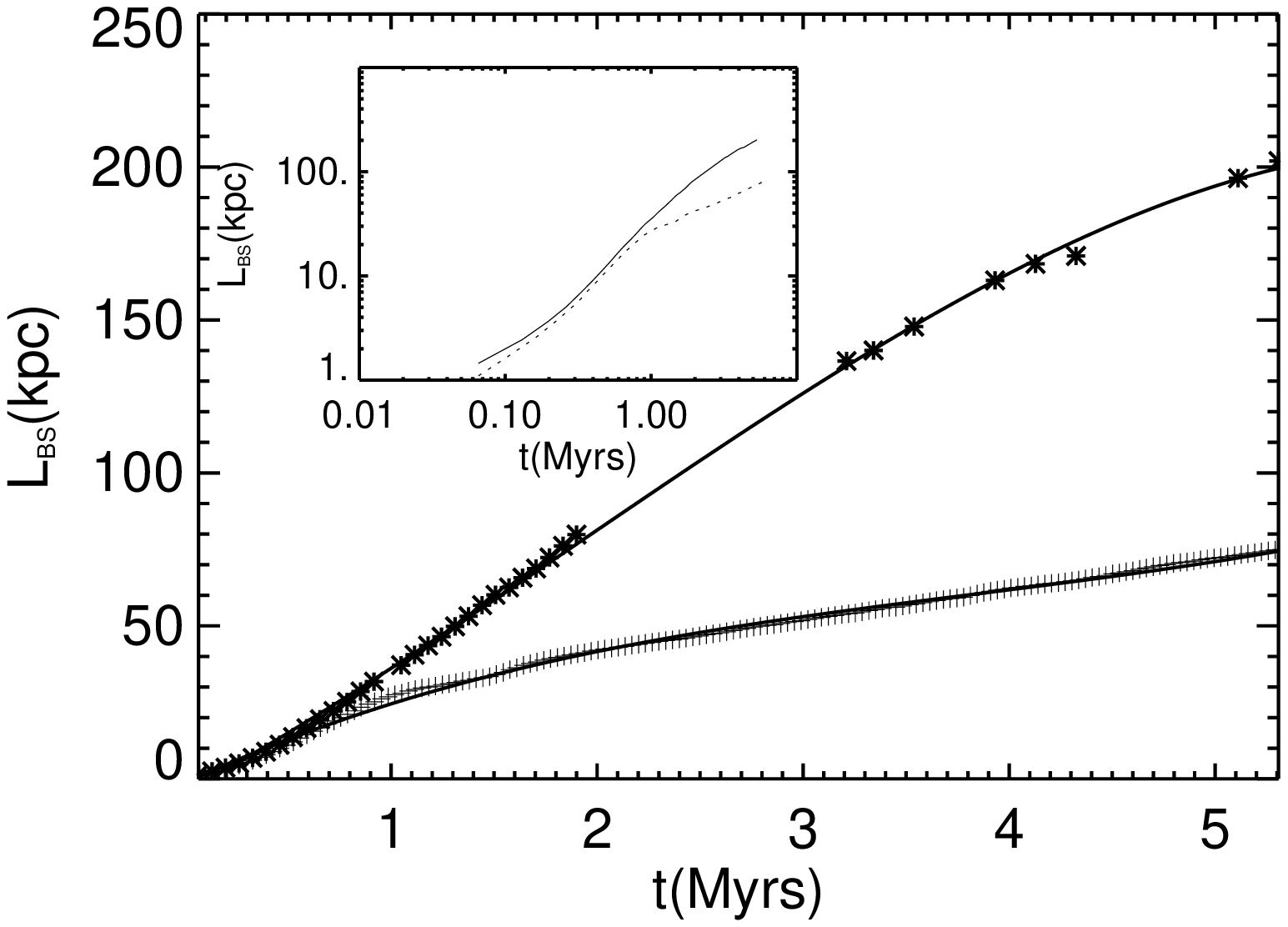} \,
\includegraphics[width=1.\columnwidth]{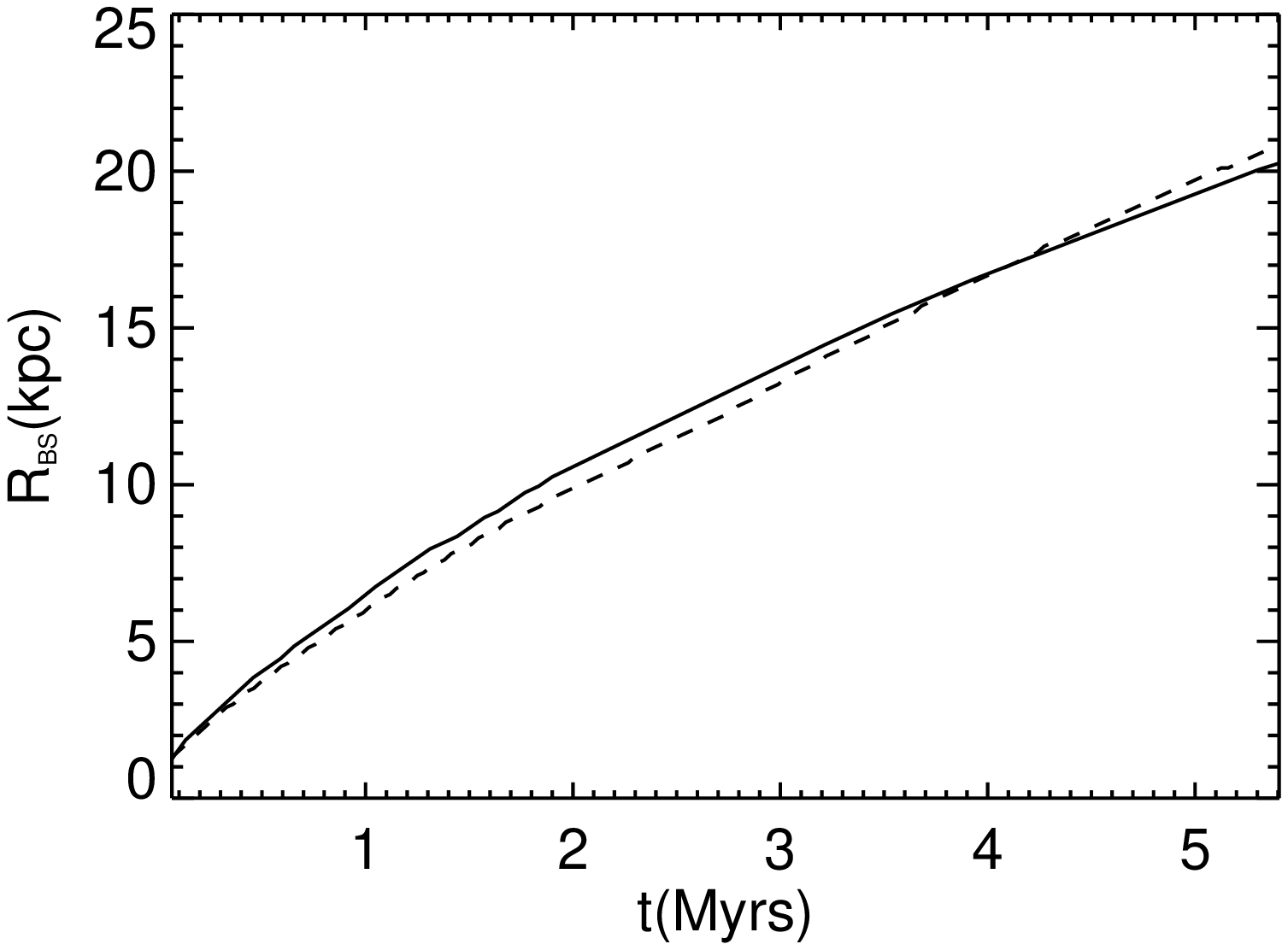}\\
\includegraphics[width=1.\columnwidth]{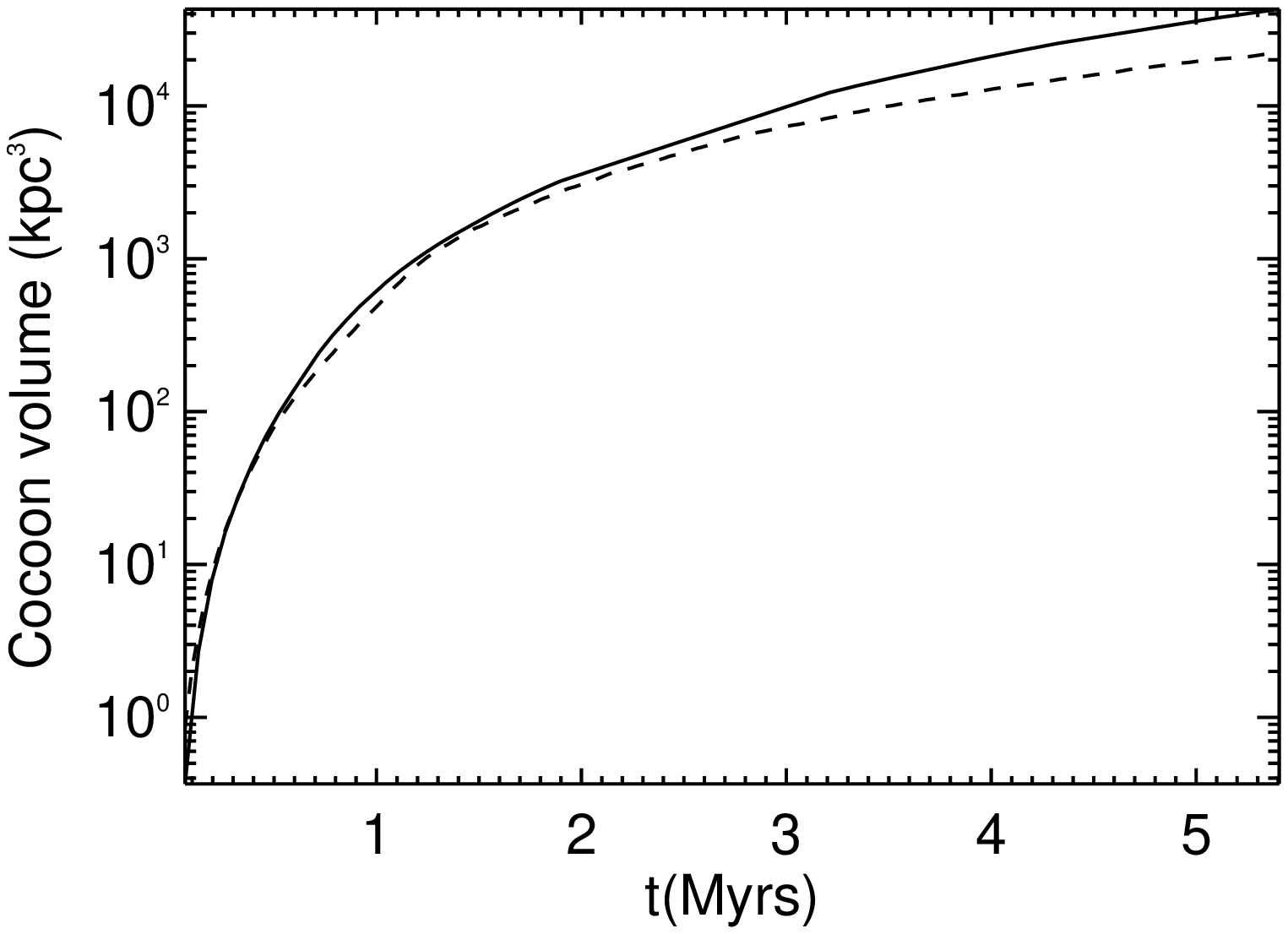} \,
\includegraphics[width=1.\columnwidth]{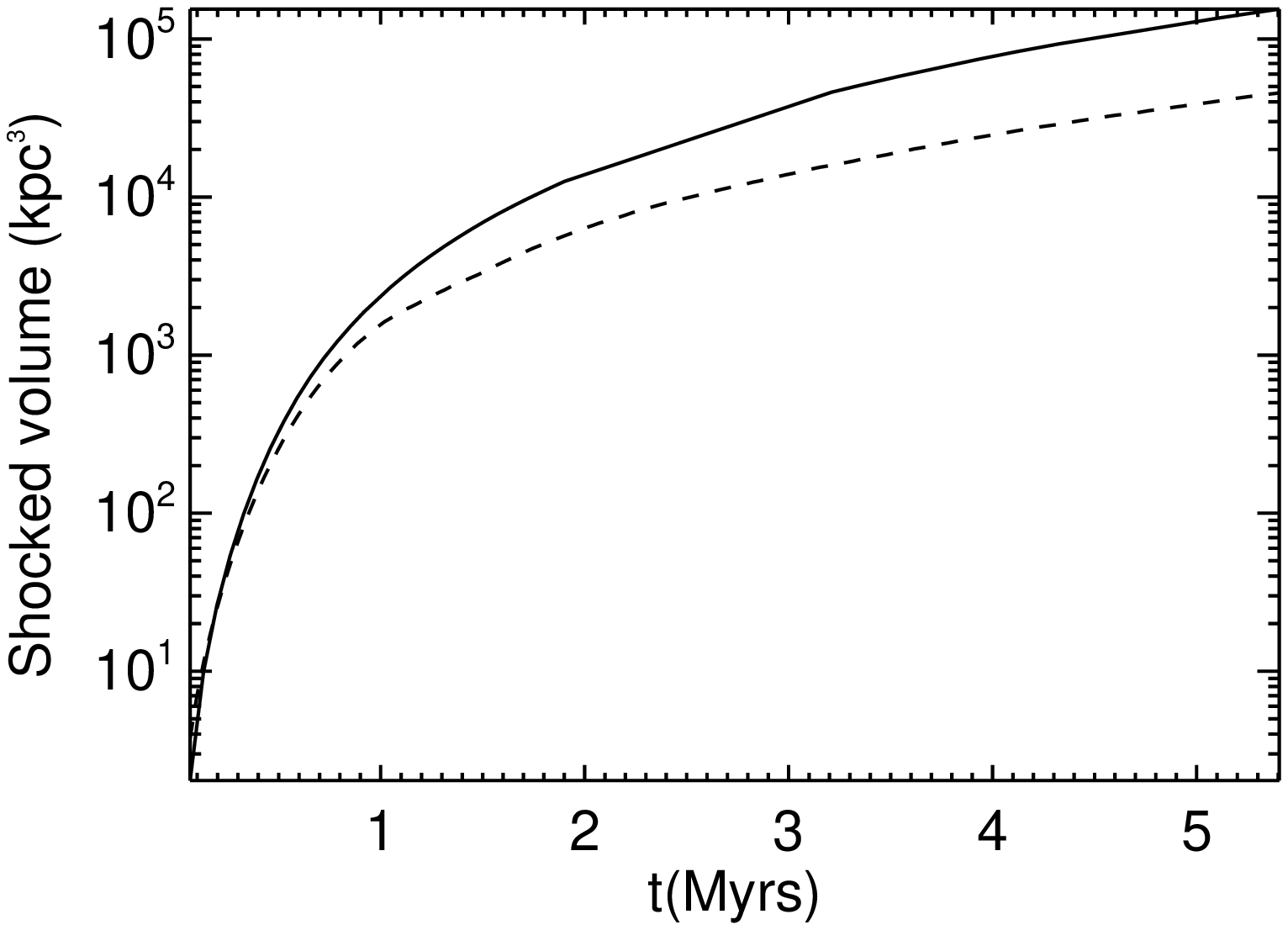}
\caption{Time evolution of the bow-shock tip position (top left panel), bow-shock maximum radius (top right), cocoon volume (bottom left), and shocked volume (bottom right) for the time span of simulation J0. In the top left panel, the asterisks indicate the 3D simulation, the crosses indicate the 2D case, and  the solid line is a third order polynomial fit that shows the accelerating and decelerating phases; the inlet shows the log-log version of the plot, with the solid (dashed) line representing the 3D (2D) simulation. In the other panels, the solid (dashed) line stands for the 3D (2D axisymmetric) simulation.}
\label{fig:kine}
\end{figure*}
%
  
Figure~\ref{fig:inner} shows a comparison of the jet head (tip of the bow-shock) position with time between J0 and several two- and three-dimensional simulations of the early jet evolution ($0-2$ Myrs). The top panel displays the head position of model J0 and several 2D-axisymmetric simulations with a resolution of 1 (J45l), 2, 4 and 8 cells per beam radius. The central panel compares the jet head position of model J0 with model J0b (J0 with a decreased initial perturbation amplitude) and again J45l. The bottom panel of Figure~\ref{fig:inner} shows the comparison of the 3D simulations using different resolutions. There are a number of relevant results to be highlighted:

\begin{itemize}

\item
The top panel of Fig.~\ref{fig:inner} shows a clearly different behaviour between the 3D simulation and the axisymmetric ones (including J45l) beyond the first 0.7 Myrs and reflects qualitative differences between 3D and axially-symmetric simulations in the dynamical processes leading to the initial evolution of the jet/cocoon system. 

\item
The properties of the perturbations triggered at the jet base clearly modify the development of these 3D effects. As seen in the central panel of Fig.~\ref{fig:inner}, the early jet evolution of simulation J0b, with much smaller initial perturbations than J0, follows that of the axisymmetric one, J45l. Thus, we can relate the differences between the 2D and 3D simulations as caused by the extra degrees of freedom in 3D models, which allows the otherwise normal jet's terminal shock to {\it wobble} around the jet axis reducing the deceleration efficiency of the jet flow \citep{aloy99}.

\item
The bottom panel shows that a larger resolution in 3D favours the wobbling of the jet's terminal shock and results in a faster advance. On the one hand, the increase in the numerical resolution improves the numerical representation of the shock and its motions. On the other hand, improved resolution allows for a faster growth of the instability amplitude \citep[see, e.g.][]{pe04}, which contributes to enhance the effect, during the linear phase. Convergence for this phase is reached at 4 cells per beam radius. 
\end{itemize}

Regarding the faster advance of the jet in the 3D simulation at this early phase of evolution, a recent work by \cite{ro17} also finds enhanced jet acceleration for the case of a jet propagating along a pressure decreasing atmosphere and a helical perturbation set at the jet base, although the simulations are restricted to the inner $\sim 10\,{\rm kpc}$. In their case, this effect is observed precisely in the only jet for which the perturbation does not reach a large amplitude (model A in the paper), consistently with the expected growth rates of instabilities in relativistic jets \citep[see, e.g.,][]{pe10}. The jets that develop large amplitude perturbations are decelerated by the dentist drill effect \citep{sch74}. In \cite{ro17}, the authors relate the jet's head acceleration to the fall in the ambient density. In this work, we show that the growing instability also contributes to this effect.

The early evolution of model J0 shows that small 3D effects influence (increase) the jet propagation efficiency. But at the same time, these effects (and consequently the characteristics and duration of the initial phase) depend strongly on the initial amplitudes and growth rates of the injected perturbations, and, implicitly, on numerical resolution. In other words, there is not a {\it unique} evolution of jets in 3D and different evolution can result for different numerical resolutions. Concerning this, let us point out that, given the large amount of numerical resources required, we were constrained to use the lowest resolution, which allowed us to reach larger physical scales and also compare with the 2D simulation with the same resolution, model J45l. We address the reader to the Appendix were we have discussed the issue of the numerical convergence of our simulations in connection with their physical credibility in more depth.

It is also relevant to note that an inhomogeneous ambient medium, with a cold component, would significantly change the propagation velocity of the jet through the inner kiloparsecs within the host galaxy \citep[see, e.g.,][]{wbu12,mu18b}. 

\subsection{Long term evolution}

%
\begin{figure*}
\includegraphics[width=\textwidth]{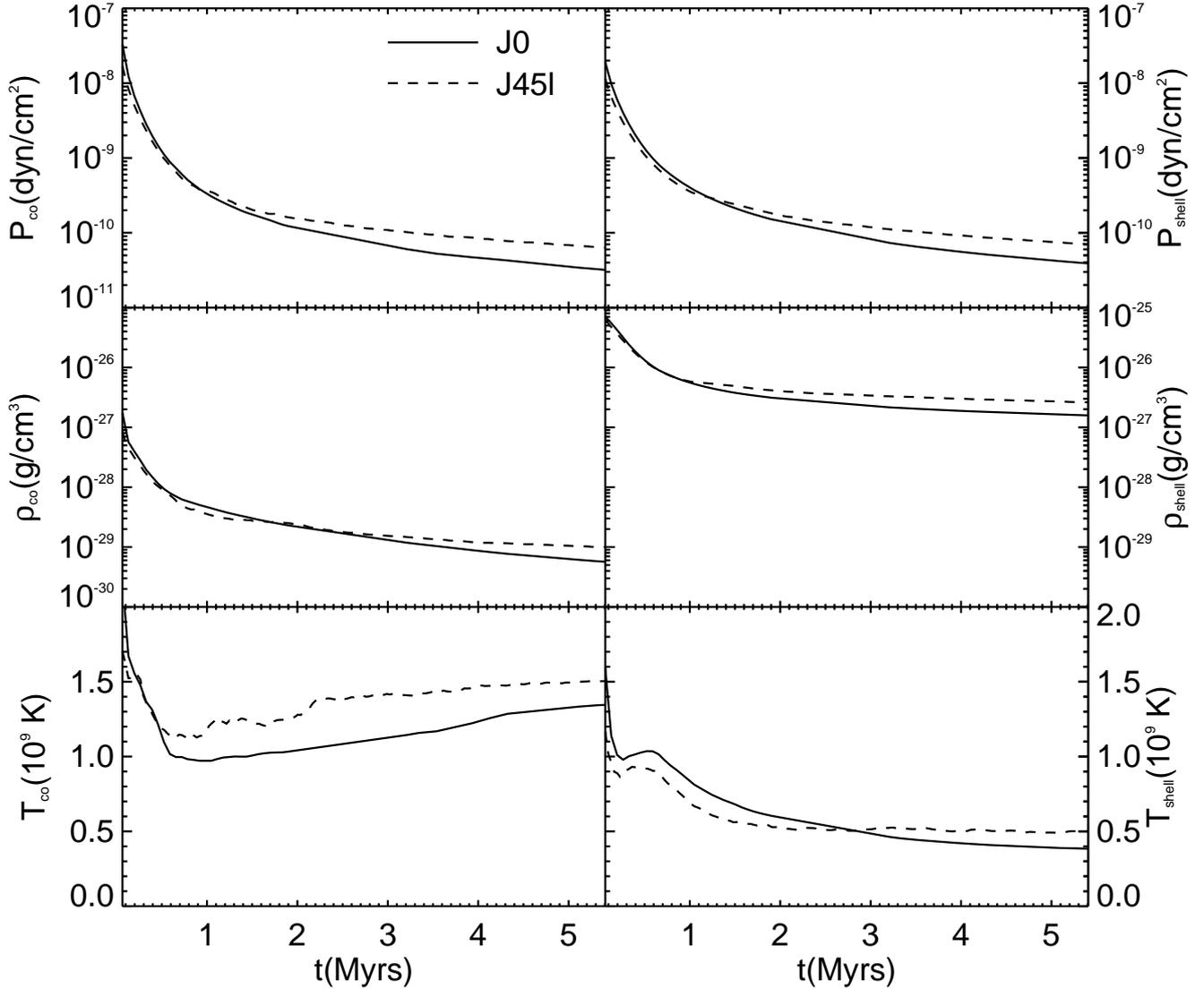}
\caption{Mean values of pressure (top), density (centre) and temperature (bottom) in the cocoon (left column) and shocked ambient medium (shell, right column). The solid line (dashed) stands for the 3D (2D axisymmetric) simulation.}
\label{fig:prhot}
\end{figure*}
%

The warnings raised in the previous section concerning the uniqueness of the jet/cocoon system during the early jet propagation should be repeated here concerning the long term evolution. This genuinely {\it multidimensional} phase is driven by a decelerating jet head as the result of the multidimensional effects affecting the jet propagation. The effects leading to this multidimensional phase refer to the interaction of the jet flow with internal (e.g., the injected perturbations) and/or cocoon-driven instabilities, which tend to make the jet advance less efficient. Most of this interaction arises in the transonic regime at the jet-backflow shear layer, prone to the development of turbulence which is, in essence, a non-linear, non-convergent process. On the contrary, the wobbling effect mentioned in the previous section relates to  the linear phase of the instability growth.

Figure~\ref{fig:kine} shows the evolution of the jet head position, the radius of the shocked region at the jet base, and the cocoon and shocked volumes for simulation J0. The shocked region is defined as the set of cells for which the initial pressure has been modified by the passage of the shock wave, which is determined by a change of a factor of 2 in the initial jet pressure. Regarding the cocoon, it is defined as the subset of cells in the shocked volume with a jet mass fraction larger than 1\% (i.e., jet tracer $f>0.01$). Finally, the shell is defined as the complementary subset. In all the panels, the evolution of the 2D axially-symmetric model J45l is also shown for comparison\footnote{From now on, the 2D axisymmetric simulation J1/J45l will be referred to as the 2D simulation, whereas J0 will be referred to as the 3D simulation.}. At the end of the 1D phase of the 2D model (0.9 Myrs), when the head of the jet starts to decelerate, its propagation speed is already smaller than that of the 3D model (see top-left panel of Fig.~\ref{fig:kine}). On the other hand, the jet head of the 3D model keeps accelerating in the density decreasing atmosphere up to 2.0 Myrs (the end of the initial phase of model J0). Beyond this time and although the rate of deceleration is the same in both simulated jets (as well shall see in Section~\ref{ss:am}), the propagation speed of the 3D model is always larger. A polynomial fit of third order of the jet head position as a function of time has been performed, showing a positive coefficient for the second degree term and a negative coefficient for the third degree one ($L_{\rm BS} = a_0\, +\, a_1 t \, + \, a_2 t^2 \, + \, a_3 t^3$, with $a_0 = -4.3\pm0.9$, $a_1 = 36.4\pm1.6$, $a_2 = 4.9\pm0.8$, $a_3 = -0.85\pm 0.10$, when $L_{\rm BS}$ is in kpc and $t$ in Myr, and where the errors are given by the poly\_fit IDL routine). This result shows that the data reproduce an axial expansion of the jet with a varying (first positive, then negative) acceleration. The break between these two behaviours occurs at 1.9 Myrs, which is the reference time we have used to separate the evolution between the initial and multidimensional phases (2 Myrs). It is important to stress that numerical viscosity slows the growth of instabilities \citep[see, e.g.,][]{pe04} and this deceleration thus occurs at smaller distances for increased resolutions. 
 
 The maximum shocked region radius (top-right panel) is similar in both 2D and 3D simulations, indicating that the radial expansion depends little on the jet advance speed (we shall come back to this point at Section~\ref{ss:am}). The volume occupied by the cocoon (bottom-left panel) is the same in the two simulations during the 1D phase of model J45l. By the end of the simulation, the volume of the cocoon in the 3D model doubles the one of the axisymmetric simulation as a result of the larger axial size of the shocked region in the 3D case. The bottom right panel shows the volume of the shocked region, typically one order of magnitude larger than the cocoon volume. At the end of simulation, the shocked volume is a factor four larger in the 3D jet.
 
 Figure~\ref{fig:prhot} shows the time evolution of the thermodynamical variables both in the cocoon and in the shocked ambient medium (shell). In the long term, the values of the pressure and density corresponding to the 2D axisymmetric simulation are larger (about a factor of 2 at the end of the 3D simulation). The temperature, on the contrary, has a very similar value in the two simulations ($\simeq 10\%$ difference in the cocoon; $\simeq 20\%$ in the shell). As it will be discussed in Section~\ref{ss:ch}, these and other results about the evolution of the thermodynamical variables in the cocoon/shell system can be explained ultimately as a consequence of the differences in the evolution of the cocoon volume between both simulations.

\begin{figure} 
\includegraphics[width=\columnwidth]{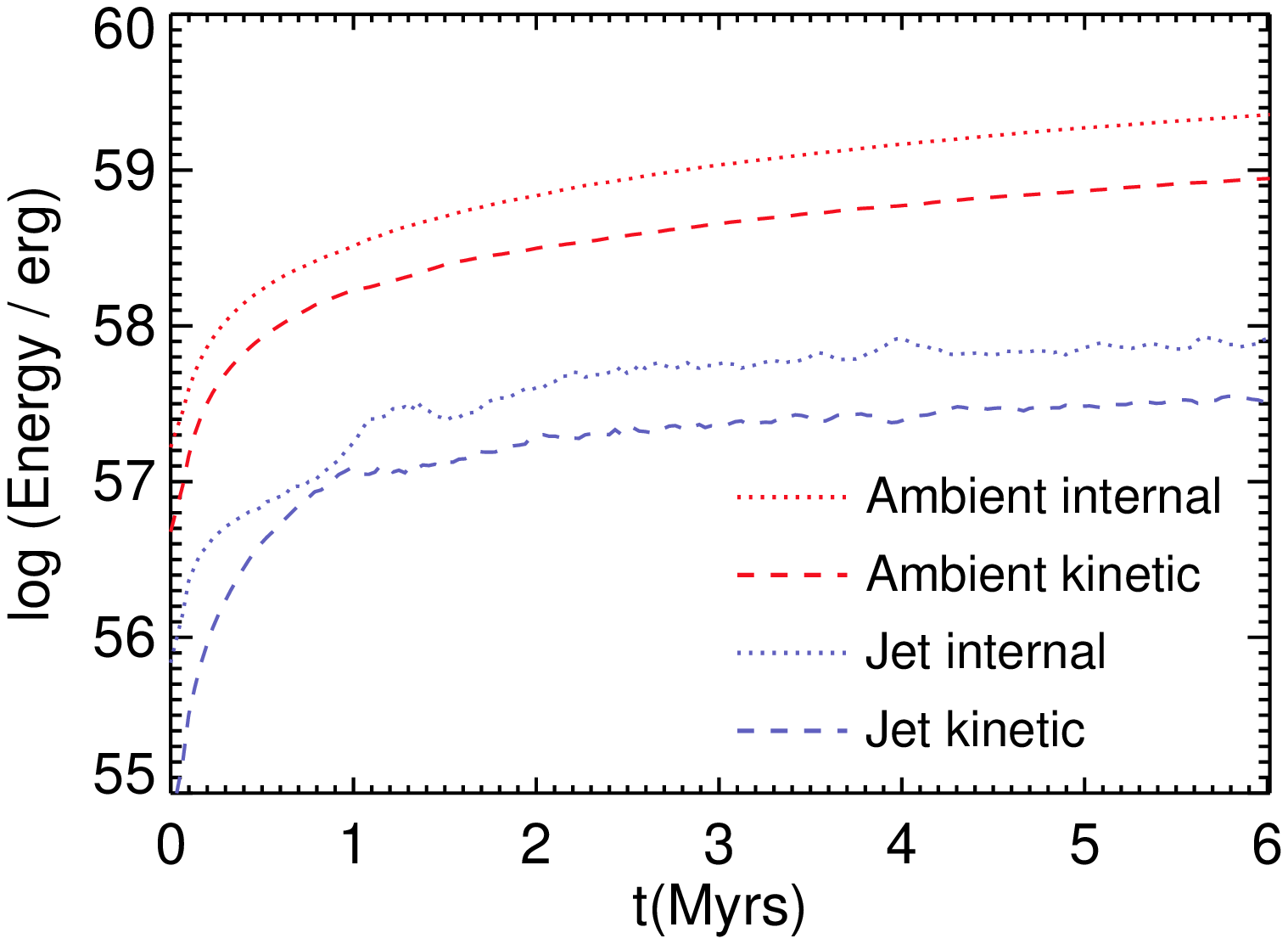}
\includegraphics[width=\columnwidth]{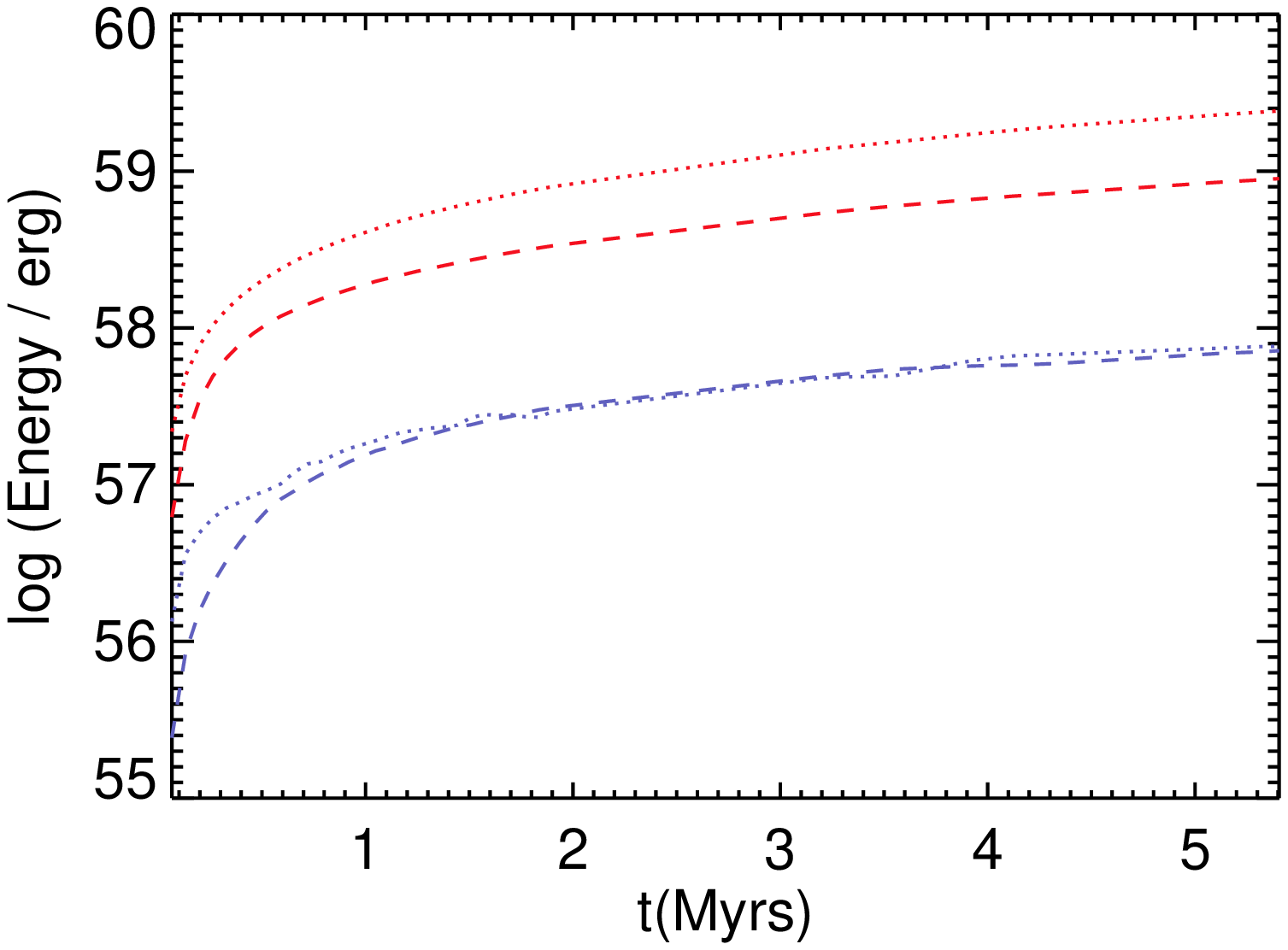}
\includegraphics[width=\columnwidth]{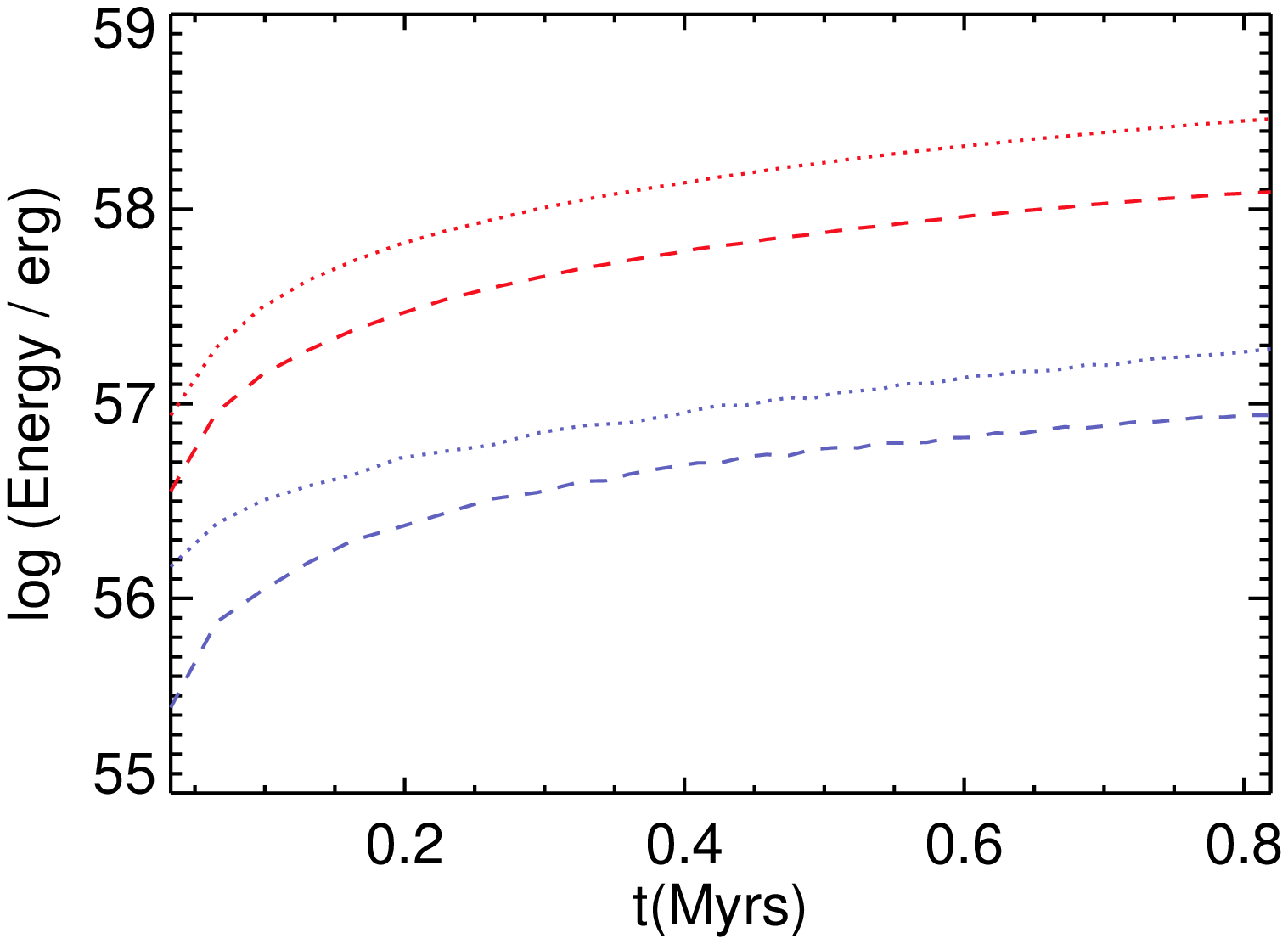}
\caption{Time evolution of the logarithm of the energy for three simulations. Top panel:  the 2D simulation labeled as J45l (PMQR14). Central panel:  simulation J0. Bottom panel: J4 simulation. The red dotted (dashed) line represents the increase of internal (kinetic) energy in the processed ambient medium. The blue dotted (dashed) line displays the internal (kinetic) energy of the shocked jet material (i.e., cocoon).}  
\label{fig:ene}
\end{figure}

 Figure~\ref{fig:ene} shows the distribution of the total injected energy in the 2D and 3D simulations along the time. The plots show the internal and kinetic energy gained by the ambient medium (red dotted and dashed lines, respectively), by substracting the original energy of the ambient gas in a given cell to that at any given time, and the internal and kinetic energy kept by the jet gas (blue dotted and dashed lines, respectively). Taking into account that the relativistic effects in the reckoning of the total energy are important only in the central jet which occupy a negligible fraction of the total volume,  the energies are computed according to the corresponding classical expressions:
\begin{equation}
\nonumber
\Delta E_{\rm int, \, a} (t) = \sum_{\rm cells} \Big(\rho(t) \, \varepsilon(t) \, \Big(1-f(t)\Big) \, - \, \rho(0) \, \varepsilon(0) \Big) \, V,
\end{equation}
\begin{equation}
\nonumber
 \Delta E_{\rm kin, \, a} (t) = \sum_{\rm cells} \frac{1}{2} \rho(t) \, v(t)^2 \, \Big(1-f(t)\Big) \, V,
\end{equation}
\begin{equation}
\nonumber
 E_{\rm int, \, j} (t) = \sum_{\rm cells} \rho(t) \, \varepsilon(t) \, f(t) \, V,
\end{equation}
\begin{equation}
\nonumber
 E_{\rm kin, \, j} (t) = \sum_{\rm cells} \frac{1}{2} \rho(t) \, v(t)^2 \, f(t) \, V,
\end{equation}
where $V$ is the volume of a numerical cell and the summation extends over all the numerical cells. The rest of the quantities were defined at Sect.~\ref{ss:code}. Since we use the classical expressions and do not account for the variation of the potential energy of the plasma along the expansion of the cavity, the sum of the different values equals the total injected energy by the jet into the grid up to a given time, $L_{\rm k} \, t$, with an error of a few percent. The comparison of the top and central panels shows that the sharing of energies is very similar in the two cases with a very efficient transfer of the injected power to the ambient medium ($\sim 80-90$\% of the injected energy), mainly in the form of internal energy, and minor differences in the relative amount of internal and kinetic energy stored in the original jet matter. The bottom panel shows that the same energy distribution is obtained for a 3D simulation with a resolution of 4 cells/$R_{\rm j}$, at the convergence level of the jet advance speed (Fig.~\ref{fig:inner}). As it can be seen at the top and middle panels of Fig.~\ref{fig:ene}, the fraction of the total injected energy that is stored in the form of ambient medium internal energy is approximately $70\%$ and $84\%$ at $t \,=\, 5.4\,{\rm Myr}$, for J45l and J0, respectively; for the ambient kinetic energy we obtain $\simeq 26\%$ and $\simeq 11\%$, whereas the energy stored by the jet particles is about $4\%$ and $5\%$. The approximate values corresponding to the control simulation J4 (bottom panel of Fig.~\ref{fig:ene}) at the end of the simulation ($0.8\,{\rm Myr}$) are, respectively, $64\%$, $29\%$ and $7\%$, which almost coincide with those corresponding to simulation J0 at the same time. In \cite{pe17}, the authors explain this efficiency on the basis of the relativistic nature of the jet causing a high hot-spot pressure, which ultimately results in an efficient energy transfer. This effect is, in contrast, not observed in mildly relativistic jets \citep[see, e.g.][]{ehk16}. 
 
 This is one of the main results of the present paper, confirming the efficiency of the heating mechanism of the ambient medium found in previous 2D axisymmetric simulations, the main difference lying on the kinematics and the region/volume in which the energy is deposited. We shall come back to these points in Section~\ref{ss:ch}.

%
%

\subsection{The large-scale picture} \label{ss:lsp}

%
\begin{figure*} 
 \includegraphics[width=0.9\textwidth]{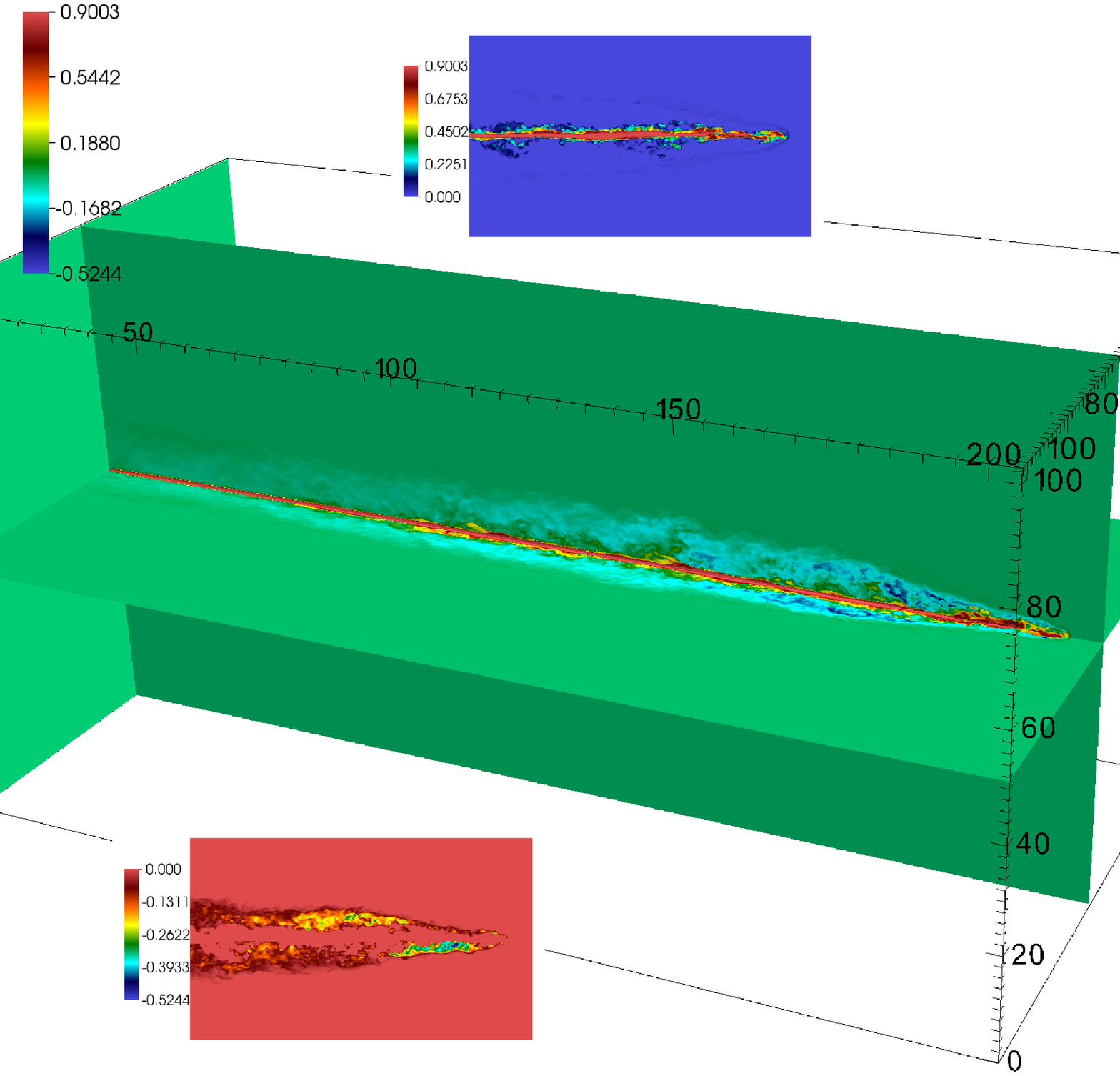}
    \caption{Cuts of the axial velocity distribution colour coded and normalized to $c=1$. The upper sub-panel shows a 2D cut from $\simeq 120\,{\rm kpc}$ to the end of the grid, displaying only the positive values of the velocity. The lower sub-panel shows the same region, for the negative values of velocity (backflow). The axial coordinates are in kpc.}  
    \label{fig:vely}
\end{figure*}
%

\begin{figure*} 
\includegraphics[width=0.9\textwidth]{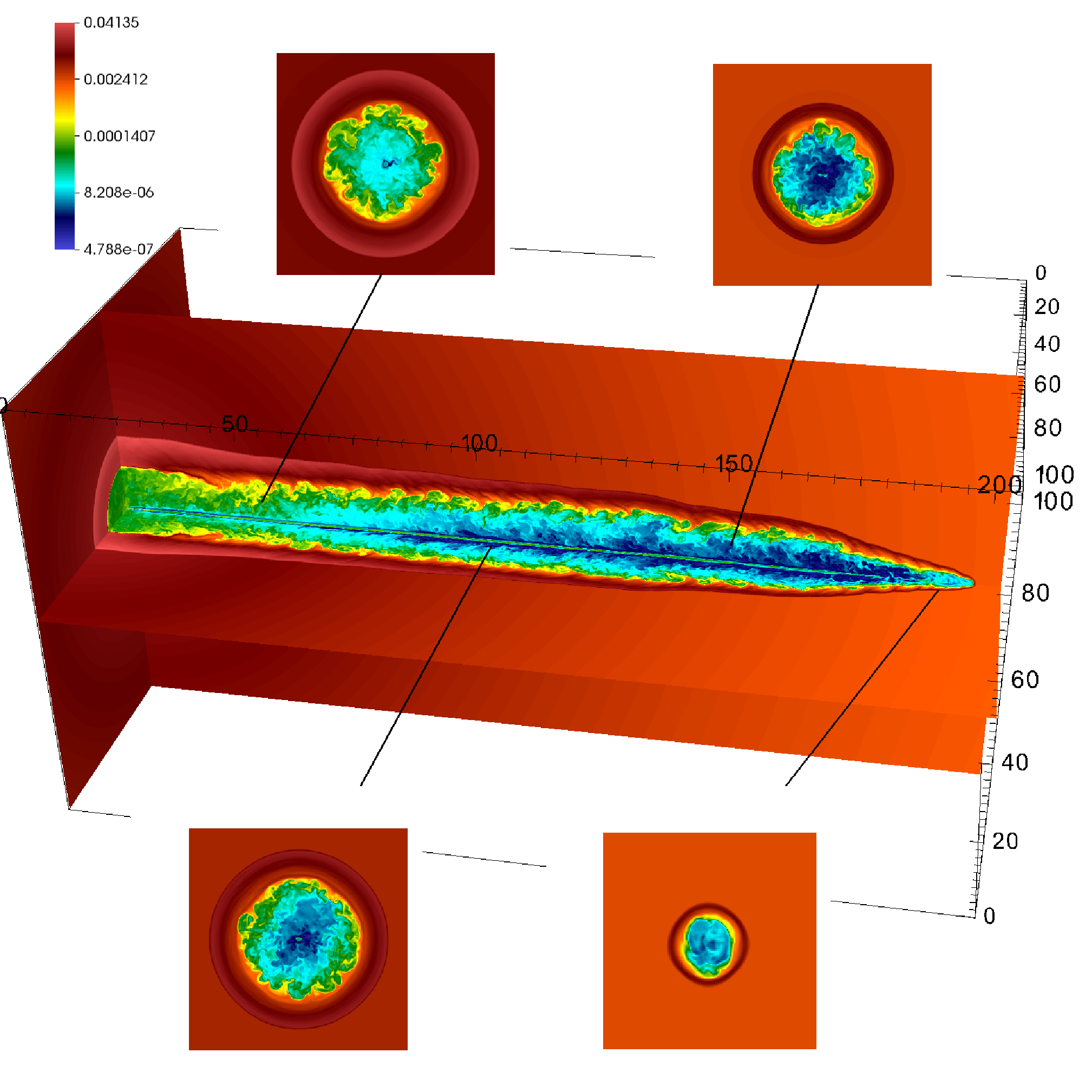}
\caption{Cuts of the logarithm of rest-mass density colour coded and normalized to the code units ($\rho_{a,0} = 1$). The sub-panels show transversal cuts at the locations indicated by the lines ($\simeq 50, \, 100, \, 150$, and $195\,{\rm kpc}$), with the same colour scale as the main image. The axial coordinates are in kpc.}  
\label{fig:densty}
\end{figure*}

\begin{figure*} 
\includegraphics[width=0.9\textwidth]{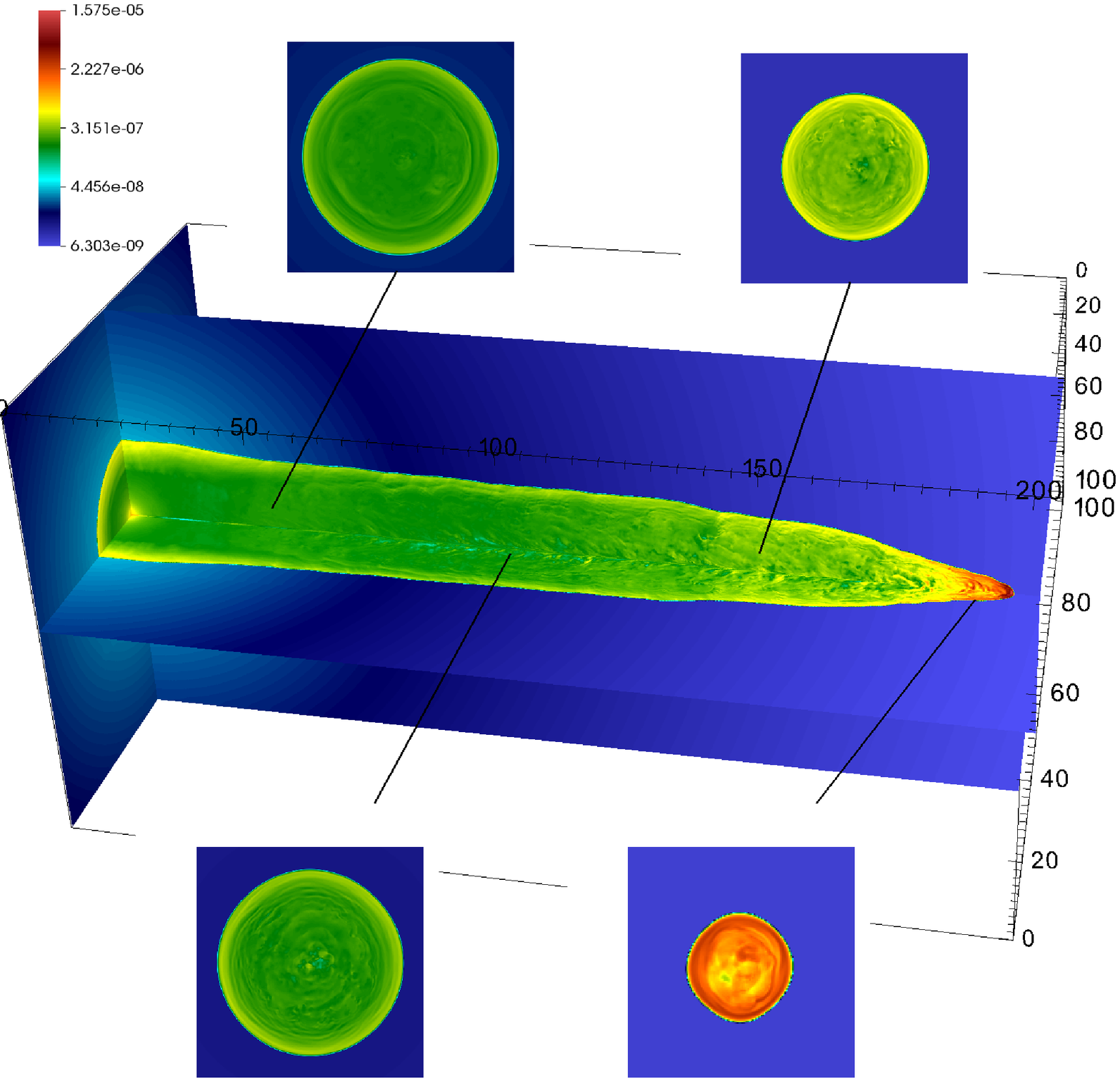}
\caption{Cuts of the logarithm of pressure colour coded and normalized to the code units ($\rho_{a,0} c^2= 1$). The sub-panels show transversal cuts at the locations indicated by the lines ($\simeq 50, \, 100, \, 150$, and $195\,{\rm kpc}$), with the same colour scale as the main image. The axial coordinates are in kpc.}  
\label{fig:pres}
\end{figure*}

Figures~\ref{fig:vely}-\ref{fig:pres} show the final snapshot of the simulation for velocity, density, and pressure, respectively. 2D transversal cuts of the jet structure at different positions along the jet axial direction are also shown. Fig.~\ref{fig:vely} shows the axial velocity. The inlet panels show only positive velocities and only negative ones (backflow), respectively. In this Figure, the jet shows an straight, collimated structure, which can be explained by the small amplitude of the helical oscillations up to $L\simeq 80~{\rm kpc}$. This is precisely the distance travelled by the jet along the first 2 Myrs (the early evolutionary phase characterized by a high jet's propagation efficiency) and coincides with the evolutionary phase in which the perturbation has linear amplitudes. Beyond this distance, although the jet is still collimated, we observe oscillations of the jet cross sections in the maps, which implies that the interaction with the ambient medium becomes harder, and this translates in jet head deceleration (see Fig.~\ref{fig:kine}). This is the onset of the \emph{dentist-drill} effect \citep{sch74}. 

%
%

The cut of the rest-mass density distribution (Fig.~\ref{fig:densty}) allows to identify the cocoon as the blue/green/yellow region surrounding the jet (green, almost straight line along the cut's symmetry axis) and the shocked ambient region (red shell wrapping the cocoon). Although typical Kelvin-Helmholtz features are observed at the contact discontinuity between the cocoon and the shocked ambient in this figure, little mixing occurs and the cocoon keeps a very low density compared to the shocked ambient medium. This region is the seed for the large cavities observed as X-ray cavities, and obtained in axisymmetric numerical simulations after the active phase (PMQR14). 
   
\begin{figure*} 
\includegraphics[width=0.75\textwidth]{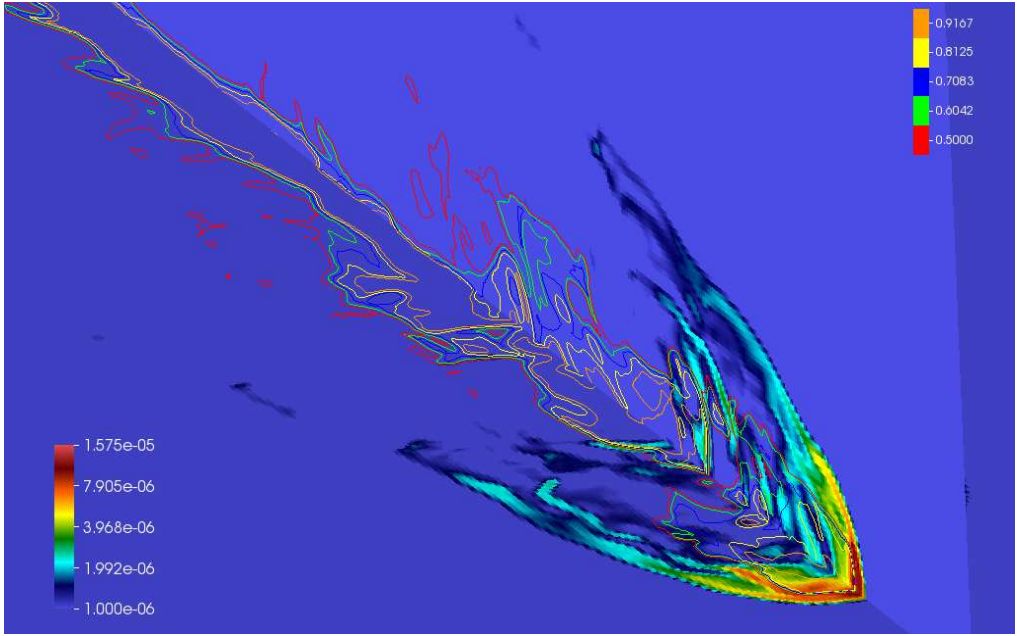}
\caption{Cut of pressure in code units ($\rho_{a,0} c^2= 1$, bottom left scale) showing the region around the jet head (hot-spot). The colour lines stand for isocontours of the tracer (jet-mass fraction, top right scale).}  
\label{fig:tracer}
\end{figure*}

    The pressure map shown in Fig.~\ref{fig:pres} shows a homogeneous shocked region, as expected from the high sound speed of the shocked gas, and a hotspot at the interaction region. The image reveals that the hotspot can be between one and two orders of magnitude overpressured with respect to the cocoon. Figure~\ref{fig:tracer} shows a zoom into the hotspot region. In this figure, we have set a minimum value of the pressure, one order of magnitude below the maximum pressure at the hotspot, which permits better visualisation of the over-pressure. The contours correspond to different jet-mass fractions, between 0.50 and 0.92. 

%
%

\section{Discussion}

\subsection{Evolution compared to different analytical models}  
\label{ss:am}

A simple description of the jet expansion can be done on the basis of the extension of the \citet{bc89} model \citep{sch02} that assumes a dependence of the jet propagation speed $v_{\rm c}$ with time in the form of a power law, $v_c \propto t^\alpha$, a complete and instantaneous conversion of the injected jet power into internal energy of the shocked gas and a sideways expansion of the (strong) shock. The model can also account for a density decreasing environment (PQM11).

%
\begin{table*}  
\begin{center}
{\small
\begin{tabular}{ll | ccccc | ccccc |}\hline
&&&{\bf Phase I}&&&&&{\bf Phase II}&&&\\
&&{$\alpha$}&{$\beta$}&{$P_c$}&{$R_{\rm BS}$}&$A_{\rm BS}$&{$\alpha$}&{$\beta$}&{$P_c$}&{$R_{\rm BS}$}&$A_{\rm BS}$\\
\hline
J45l & Sim  &   $0.07$ &  $-1.55$ & $-1.58$  &  $0.75$  & $0.32$ & $-0.23$  & $-0.52$  &  $-1.09$ & $0.66$ & $0.11$\\
   & Model&        &        & $-1.65$  &  $0.79$  & $0.28$ &       
&         &
$-1.05$ &
$0.64$ & $0.13$\\ \hline
J0 &  Sim &   $0.30$ & $-1.55$  & $-1.74$  & $0.65$ & $0.65$ & $-0.22$  &
$-0.52$  &  $-1.20$ &
$0.65$ & $0.13$ \\
   & Model&        &        & $-1.69$  & $0.69$ & $0.61$ &       
&         &
$-1.06$ &
$0.64$ & $0.14$\\  \hline
\end{tabular}
}
\caption{Values of the exponents of the power laws determining the time evolution of the cocoon pressure ($P_c$), the transversal size of the shocked region ($R_{\rm BS}$) and the shocked region aspect ratio ($A_{\rm BS}$) during the two phases of the evolution of models J45l (2D-axisymmetric) and J0 (3D). The parameters $\alpha$ and $\beta$ are derived from the simulations. The time dependence of $P_{\rm c}$, $R_{\rm BS}$, and $A_{\rm BS}$ is shown as obtained from the simulation and from the eBC model. The time separating phases I and II in the two simulations is different (0.91 Myrs for J45l; 2.0 Myrs for J0). The duration of Phase II is also different (50.0 Myrs for J45l; 5.0 Myrs for J0).}
\label{tab:tab2}
\end{center}
\end{table*}
%

The axial expansion of the jet is governed by the ram pressure equilibrium at the jet's head. In the absence of relevant multidimensional effects or changes in the injected power, the expansion proceeds at a constant speed \citep[the so-called one-dimensional propagation speed][]{ma97}, $\alpha = 0$. This propagation speed can increase during the one-dimensional phase for jets propagating through density decreasing atmospheres ($\alpha > 0$). During the multidimensional phase, the development of instabilities tend to make the jet advance less efficient and hence reduce the jet's propagation speed, $\alpha <0$. However, note that the intermittency of the processes affecting the jet's working surface (the jet/ambient medium interface at the jet's head) make the evolution of the jet propagation speed non-monotonic.

These two phases can be easily identified in axisymmetric simulations \citep[e.g.,][PQM11]{sch02,pm07} and the evolution of the axial and radial expansion, and cocoon pressure, is consistently described. In particular, for the 2D axisymmetric simulation J45l, with the same injection parameters as the 3D simulation discussed in this paper, PQM11 give a 1D phase lasting for 0.91 Myrs with $\alpha = 0.07$ (as a result of the propagation through the density decreasing atmosphere), and a multidimensional phase with $\alpha = -0.23$ up to the end of the active phase, at 50 Myrs.

The expansion of the cavities in three dimensions is more complex. In this discussion, we divide the evolution of the shocked region of model J0 still into two phases, namely initial phase and multidimensional phase (phases I and II, respectively). The initial phase corresponds to the one-dimensional phase of axisymmetric models but incorporates small three-dimensional effects within the beam that tend to make the jet propagation more efficient. In particular, a simple structure like a normal terminal Mach shock at the jet head will probably not survive when significant 3D effects within the beam develop \citep[e.g.,][]{aloy99}. A Mach shock which is no longer normal to the jet will decelerate the jet flow less efficiently, facilitating its axial expansion. However, this {\it shock wobbling} \citep[as it was dubbed in][]{aloy99} represents only a transient phase until the growth of three-dimensional perturbations force the beam to spread its momentum over a much larger area, reducing the jet advance speed drastically. It is also possible that the so-called multidimensional phase could be divided into  a {\it post-linear} three-dimensional phase  and a {\it non-linear phase}, much more effective in head deceleration \citep[as described in the {\it dentist-drill model,}][]{sch74}.

We thus consider the evolution of simulation J0 through an initial phase of 2.0 Myrs, the values of $\alpha$ taken from a log-log fit of the bow-shock tip position as a function of time, and the eBC model presented in PQM11. In this model the advance speed of the bow shock along the axial direction, $v_{\rm c}$, and the ambient density, $\rho_{\rm a}$, follow the power laws $v_{\rm c} \propto t^{\alpha}$, $\rho_{\rm a} \propto r^\beta$. Parameter $\alpha$ controls the axial expansion rate under both the internal beam processes affecting the jet head propagation, and the density decreasing environment. Parameter $\beta$ regulates the sideways expansion. According to the eBC model, the transversal dimension of the shocked region, $R_{\rm BS}$ (with subscript $BS$ referring to the bow shock), and cocoon pressure, $P_{\rm c}$, follow

\begin{equation}
\label{eq:eBC}
\displaystyle{R_{\rm BS} \propto t^{\frac{2 - \alpha}{4 + \beta}}}, \,\,
\displaystyle{P_{\rm c} \propto t^{\frac{2(\alpha - 2) - \alpha(4 +
\beta)}{4 +
\beta}}}.
\end{equation}

Table~\ref{tab:tab2} shows the values of the exponents of the power laws for the jet head propagation speed, the transversal size of the shocked region, and its pressure, as a function of time for model J0 as obtained from the simulation and from the eBC model. The values corresponding to the axisymmetric simulation J1/J45l are also shown for reference.

The first conclusion is that the accuracy of the eBC model in matching the fitted slopes of the J0 simulation is good although poorer than for the axisymmetric simulation. Moreover, the fact that the values of $\alpha$ for the two phases are so different supports the initial hypothesis of dividing the overall long term evolution in two stages.

Comparing the values of $\alpha$ of the 2D and the 3D simulations for phases I and II we conclude the following: (i) the {\it shock wobbling} effect mentioned earlier (non-existent in axisymmetric models) counterbalances during a short period of time, and delays, the deceleration driven by early multidimensional effects, and allows the jet to accelerate efficiently in the density decreasing atmosphere along Phase I; (ii) along Phase II, the jet in the 3D simulation decelerates at the same rate as the 2D axisymmetric simulation, although the model fit of the pressure evolution is poorer than in the 2D case. The similarity of the evolution of both simulations along Phase II is probably an indication that the 3D simulation was too short as to capture the full three-dimensional phase. However, we point out that our description is phenomenological since the value of $\alpha$ in the different phases and the duration of the phases themselves depend on the initial amplitudes of the injected perturbations and the growth rates of the excited modes.

Finally, let us discuss the consistency of the eBC model in describing the evolution in the 3D simulation. For values of $\beta > -2$ (as in the present case), Eq.~\ref{eq:eBC} tells us that smaller values of $\alpha$ lead to faster transversal expansions and a slower decrease in pressure. Within Phase I, the increase in the value of $\alpha$ from the 2D model to the 3D one is certainly associated with a slower transversal expansion and a faster decrease of pressure. However during Phase II, the model fails to capture the steeper cocoon/shocked region pressure decrease of the 3D model. This shortcoming can have several explanations. One is that the eBC model is too simple to capture the wealth of three-dimensional processes taking place in the turbulent plasma filling the cocoon. As an example, let us remind that one of the hypothesis behind the eBC model is the randomization and homogenization of the injected energy into the whole shocked region, which can not be instantaneous in the case of a large volume in rapid expansion (see, Fig.~\ref{fig:pres}). Other possibilities are related with the limitations ($R_{\rm BS}$ is represented only by the radius at the base of the jet, Phase II is possibly too short) of the procedure to derive the power laws representing the
transversal expansion and the pressure decrease. However and despite all its limitations, the eBC model still offers a simple and consistent characterization of the expansion of the shock-driven cavities also in 3D.

%
%

Quantity $A_{\rm BS}$ in Table~\ref{tab:tab2} is the aspect ratio of the shocked region, $A_{\rm BS} = L_{\rm BS}/R_{\rm BS}$, where $L_{\rm BS}\propto v_{\rm c} \, t \propto t^{1 + \alpha}$ is the axial length. According to the values shown in the table, during Phase I the shock stretches along the jet direction becoming more and more elongated. However, once the evolution enters into the multidimensional phase the shock expands almost self-similarly, $A_{\rm BS} \propto t^{0.13}$ (strictly speaking, a self-similar expansion implies $A_{\rm BS}$ constant, $A_{\rm BS} \propto t^0$).

Theoretical as well as numerical modelling of powerful jets in density decreasing atmospheres \citep{fa91,ka97,kf98}  point to a self-similar expansion of the shock for long enough time scales (large enough spatial scales). The small axial stretching undergone by the shocks of models J45l and J0 along Phase II can be readily explained by the radial profile of the ambient density, which is described by three broken power laws, $\rho_a \propto r^\beta$, with $\beta \simeq -1.55, -0.52, -1.02$ for $r <10$ kpc, $10$ kpc $< r < 100$ kpc, and beyond 100 kpc, respectively. Along Phase II the sideways expansion of the shock proceeds across the middle region with the flatter ambient density profile ($\beta = -0.52$) while the head of the jet has already reached the outer region with a steeper density gradient ($\beta = -1.02$). The greater ease of the shock expansion along the axial direction gives rise to the relative excess of axial stretching.

Nevertheless, we recall that the evolution along Phase II does not correspond to the expected asymptotic long-term evolution of model J0, and this quasi-self-similar stage must be understood as a temporary phase before the triggering and development of the fully-3D perturbations that affect the jet propagation.

\subsection{Cavity's thermodynamical evolution and transfer of energy to the ambient medium}  
\label{ss:ch}

In Figure~\ref{fig:prhot} we showed the time evolution of the thermodynamical variables both in the cocoon and in the shocked ambient medium. Let us first concentrate on the evolution of the cocoon quantities displayed on the left column panels. According to the analytical model described in the previous section, a complete and instantaneous conversion of the injected jet power into internal energy makes the pressure in the cocoon inversely proportional to its volume, $P_{\rm c} \propto V_{\rm c}^{-1}$, the proportionality constant depending only on the jet injection conditions (which are the same in the 2D and 3D simulations). This explains the time evolution of the cocoon pressure of the 2D and 3D simulations displayed in the top-left panel of Fig.~\ref{fig:prhot} and the factor $\sim 2$ between both pressures at the end of the simulation. The time evolution of the cocoon density admits a similar explanation. 

 The density in the cocoon at any time can be readily estimated as the total mass injected into the cocoon divided by the cocoon's volume, $\rho_{\rm c} \propto V_{\rm c}^{-1}$, the proportionality constant depending again only on the jet injection conditions. This inverse proportionality between mass density and volume at the cocoon explain the profiles shown in the middle left panel and, again, the difference between the cocoon densities in the 2D and 3D models by the end of the simulation. The temperature in the cocoon, understood as a measure of the internal energy per unit mass in this region, is hence directly proportional to the pressure and inversely proportional to the mass density making it roughly constant with time (and between the two simulations), $T_{\rm c} \propto P_{\rm c}/\rho_{\rm c} \propto 1$. This is precisely the result shown in the bottom left panel of Fig.~3, which displays a variation of the cocoon temperature for the 3D simulation of $\simeq 35\%$ along Phase II and a difference of $\simeq 10\%$ with respect to the 2D model at the end of the simulation.

The evolution of the thermodynamical variables of the cavity's shell (see right panels of Fig.~\ref{fig:prhot}) is also well understood within this model. The pressure is the same as in the cocoon, $P_{\rm s} \sim P_{\rm c}$, in both simulations since the sound speed in the cocoon/shocked-ambient-medium region is about one or two orders of magnitude larger than its expansion velocity, hence allowing for an almost instantaneous (with respect  to  the  dynamical time-scale) adjustment of the pressure.\footnote{This is not exactly the case close to the hot spot (see Fig.~\ref{fig:pres}), because of the difference between the expansion speed in the axial direction and the sound speed, which still cause a pressure gradient from the jet head towards the main cocoon volume.} The density in the shell is smaller in the 3D case since at any given time it expands against a lighter medium.

One of the central points of the present work was to probe in 3D the efficiency of the heating mechanism of the ambient medium found in previous 2D axisymmetric simulations. The results at the end of sec.~\ref{ss:lsp} (see Fig.~\ref{fig:ene}) corroborate that the sharing of energies is very similar in the 2D and 3D cases with a very efficient transfer of the injected power to the ambient medium, mainly in the form of internal energy. The fact that most of the energy accumulated in the shocked ambient medium is in the form of internal energy is a consequence of the strength of the bow shock. In addition, our model of expansion of the shocked volume/cocoon offers a simple explanation of the agreement between the 2D and 3D results independently of the shape and size of the cavity, assuming a self-similar expansion for the cocoon/shocked-ambient-medium complex. Indeed, in the case where most of the energy content of the shell is in the form of internal energy, it can be estimated to be the product of the pressure at the shell (equal to the pressure at the cocoon) times the shell volume, $E_{\rm int, \, s} \propto P_{\rm s} V_{\rm s} \propto P_{\rm c} V_{\rm s}$. The cocoon pressure, according to our model, is proportional to the ratio of the injected power over the cocoon volume, and consequently, $E_{\rm int, \, s} \propto V_{\rm s}/V_{\rm c}$. Under conditions of self-similar expansion, the time dependence of both the shell and cocoon volumes cancel and the internal energy at the shell is simply proportional to the jet power, which is the same in both 2D and 3D simulations. Fitting the ratio $V_{\rm s}/V_{\rm c}$ to a power law of time, $V_{\rm s}/V_{\rm c} \propto t ^{\delta}$, we find $\delta \simeq -0.12, -0.11$ for the 2D and 3D simulations along Phase II. The mildness of the time dependence of $V_{\rm s}/V_{\rm c}$ (and of the cavity's aspect ratio discussed earlier) validates the hypothesis of self-similarity and gives consistency to our reasoning.

\subsection{Comparison with FRII sources from the 3C Catalogue}

\begin{figure} 
\includegraphics[width=0.95\columnwidth]{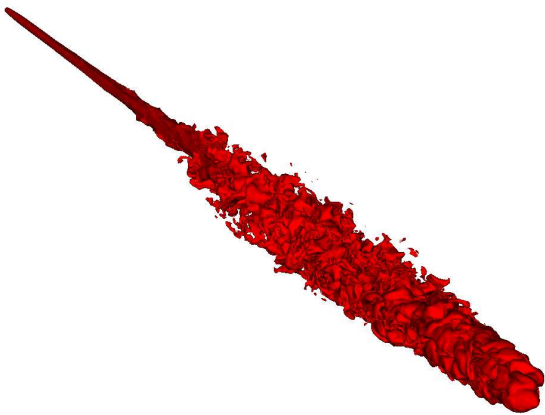}
\caption{Isosurface image of the tracer at $f=0.5$.}  
\label{fig:f05}
\end{figure}

 Taking into account that the simulated region that corresponds to radio lobes is the cocoon, we show in Fig.~\ref{fig:f05} an isosurface image of the tracer at $f=0.5$. The isosurface thus embeds the cells in which the jet mass fraction is above 50\%, precisely the region from which we would expect the synchrotron emission. The image shows a very similar morphology to that observed in many classical FRII radio galaxies \citep[e.g., 3C 33, 3C 46, 3C 219, 3C 223, 3C 228, 3C 244.1, 3C 273, 3C 274.1, 3C 321, 3C 427.1, 3C 452, 3C 457 in http://www.jb.man.ac.uk/atlas/index.html,][]{lrl83,gi04,orr10}, or giant radio-galaxies like, e.g., IGR J14488-4008 \citep{mol14}. 

%
%

\subsubsection{Lobe pressures and volumes}

%
\begin{table*}
  \begin{center}
 {\small
  \begin{tabular}{l | ccccccccccc}\hline
Source & $L_{\rm k}\,(10^{45}{\rm erg/s})$ & $T\,({\rm Myr})$& $L_{\rm BS}\,({\rm kpc})$ & $\rho_{\rm ICM}\,(10^{-28}\,{\rm g/cm^3})$ & $P_{\rm ICM}\, (10^{-12}\,{\rm dyn/cm^2})$ &  $T_{\rm ICM}\,({\rm keV})$ \\
\hline
  J0 & 1. &  5.4 & 200  & $4$ & $1$ & 1.5 \\
  \hline
  3C 33 & 0.3 & -  & $120-150$ & - & $0.09-0.14$& 1.1  \\
  3C 46 & 3.2 & - & 586 & - & $1.32$ & 2.11 \\
  3C219 & 1.8 & - & 285 & - &  $2.5$ & 1.46 \\
  3C223 & -  & 75 & $350 - 400$ & - & $0.96$ & - \\
  3C228 & 21 & - & 170 & - & $4$ & 2.2  \\
  3C244.1& 3.5 & - & 166 & - & $9.5$ & 2.05 \\
  3C267 & 10 & 20 & 170 & $10$ &  - & - \\
  3C268.1 & $5 - 11$ & $20-27$ & $170 - 200$ & $3.3 - 5$ & - & - \\
  3C274.1 & 5.4 & - & 485 & - & $0.2$ & 0.95  \\
  3C321 & 0.27 & - & 275 & - & $0.03$ & 0.87  \\
  3C322 & 25 & 12 & $140-180$ & $5$ & - & - \\
  3C427.1 & 8.4 & - & 100 & -  & $43$  & 3.14 \\
  3C427.1 & $1.4 - 1.8$ & $45 - 50$ & 80 & $40$ & - &-\\
  3C437 & $30 - 76$ & $7 - 11$ &  165 & $1.7 - 3.4$&  - &-\\
  3C452 & $0.65 - 0.76$ & $80-90$ & 210 & - & $0.7-1.1$ & 1.32 \\
  3C457 & 3.4 & - & 633 & - & $0.95$ & 3.1 \\
  Cygnus A & $6-8$ & 20 & $80-90$ & $83$ & $100$ & $6-8$ \\
\hline
  \end{tabular}
  }
   \caption{Cocoon parameters in J0 and lobe parameters for different sources with similar morphologies to the simulated jet, as given in the literature. Column 1: Source name. Col. 2: Jet power. Col. 3: Jet age. Col. 4: Linear size. Col. 5: ICM density close to $L_{\rm BS}$. Col. 6: ICM pressure close to $L_{\rm BS}$. Col. 7: ICM temperature. In sources with similar values for both lobes we have given a mean value and we have given the two values when either there were significant differences between different papers, or both lobes had significantly different values. For reference 1 \citep{od09}, we have chosen the data obtained by taking a magnetic field $B=0.25\,B_{\rm eq}$. In the case of 3C427.1, we have decided to show the parameters given by references 1 and 4 separately.}
 \label{tab:tab3}
 \end{center}
\end{table*}
%

%
\begin{table*}
  \begin{center}
 {\small
  \begin{tabular}{l | ccccccccccc}\hline
Source &  $P_{\rm l}\, (10^{-11}\,{\rm dyn/cm^2})$ & $V_{\rm l}\,(10^5\,{\rm kpc^3})$ & $v_{\rm hs}\,(\rm{10^{-2} c})$ & $M$ & Ref. \\
\hline
  J0 &  $5$ & $0.4$ & 12 & $6.3^{(*)}$ & - \\
  \hline
  3C 33 &$0.2$ & $9$ & - & $3.5-4$ & 4 \\
  3C 46 & $0.3$ & $200$ & - & 1.4 & 4 \\
  3C219 & $0.6-0.8$ & $55$ & - & $1.5-1.9$ & 4\\
  3C223 & $0.1$&  $55$ & 1.5 &  - & 2, 3\\
  3C228 &  $3.6$& $100$ & - & 2.7 & 4 \\
  3C244.1&  $3.6$ & $3.6$ & - & 1.8 & 4 \\
  3C267 &  $80 - 110$ & - & 0.03 & - & 1 \\
  3C268.1 &  $26 - 40$ & - & 2.5 - 4 & - & 1\\
  3C274.1 & $0.2$ &  $330$  & - & 2.8 & 4 \\
  3C321 & $0.6-1.8$&  $0.2-2.3$ & - & $13-21$ & 4 \\
  3C322 &  $50-110$ & - &  $3.6 - 5.0$ & - & 1 \\
  3C427.1 & $5.3$ & $3.6$ & - & 1.1 & 4\\
  3C427.1 &  $27-44$ & - & 1 & - & 1\\
  3C437 & $50-100$ & - & $5 - 8$ & - & 1 \\
  3C452 &$0.3-0.4$& $38$ & $0.7-0.8$ & 1.8 & 2, 3, 4\\
  3C457 & $0.12$& $520$ & - & 1.1 & 4\\
  Cygnus A &  $50-90$ & $4$& $1 - 2$ & $1.2 - 1.6$ & 5\\
\hline
  \end{tabular}
  }
   \contcaption{Cocoon parameters in J0 and lobe parameters for different sources with similar morphologies to the simulated jet, as given in the literature. Column 1: Source name. Col. 2: Lobe pressure. Col. 3: Lobe volume. Col. 4: Advance speed. Col. 5: Shock Mach number. Col. 6: References \citep[1-5:][respectively]{od09, har16,har17,ine17,sni18}. In sources with similar values for both lobes we have given a mean value and we have given the two values when either there were significant differences between different papers, or both lobes had significantly different values. For reference 1 \citep{od09}, we have chosen the data obtained by taking a magnetic field $B=0.25\,B_{\rm eq}$. In the case of 3C427.1, we have decided to show the parameters given by references 1 and 4 separately. $^{(*)}$: $M=50$ if computed at the head of the jet.}
 \end{center}
\end{table*}
%

  During the last decade, different groups \citep[e.g.,][]{od09, har16,har17,ine17} have provided estimates of different lobe parameters for a number of FRII sources \citep[see also, e.g.,][]{wdw97a,wdw97b}. The lobe pressure was computed using typical assumptions such as equipartition, minimum energy, or using inverse Compton (IC) together with the observed X-ray flux. Table~\ref{tab:tab3} includes a list of physical parameters of the lobes in different sources that are either similar in morphology to J0, or of similar linear size. For the sources with similar morphologies to J0, \cite{ine17} find IC lobe pressures $\sim 10^{-12}-10^{-11}\,{\rm dyn\,cm^{-2}}$ for jets with $L_{\rm BS} \simeq 100 - 300\,{\rm kpc}$ (e.g., 3C 33, 3C 219, 3C 321, 3C 452 and 3C 427.1). The authors obtain, for the same FRIIs, ICM pressures in the range $\sim 10^{-14}-10^{-12}\,{\rm dyn\,cm^{-2}}$ by the tip of the lobes.
  
  \citet{od09} give lobe pressures $\sim 10^{-10}-10^{-9}\,{\rm dyn\,cm^{-2}}$ for powerful FRII sources (typically $L_{\rm k} \sim 10^{45-46}\,{\rm erg/s}$) between 100 and 300~kpc. For the ICM, the authors provide estimates of the particle number density, between $10^{-4}$ and $10^{-3}\,{\rm cm^{-3}}$, or $10^{-6}$ and $10^{-5}\,{\rm cm^{-3}}$ at 100~kpc, depending on the magnetic field considered with respect to the minimum energy value ($B/B_{\rm eq}=0.25$ and $1$, respectively). 
    
  In our simulation, the ICM pressure at $100-200\,{\rm kpc}$ from the active nucleus is $\sim 10^{-12}\,{\rm dyn\,cm^{-2}}$ (see Figure 1 in PMQR14 and Eqs.~\ref{next} and \ref{pext}), of the order of magnitude of  the reported numbers by \cite{ine17}, but typically larger (except in the case of 3C~219, with $\sim 2.4\times10^{-12}\,{\rm dyn\,cm^{-2}}$ at $\simeq 300\,{\rm kpc}$). Regarding the ICM particle density at $100\,{\rm kpc}$, it is $\simeq 6\times10^{-4}$ cm$^{-3}$ (Figure 1 in PQMR14 and Eq.~\ref{next}), in the range given by \cite{od09} for a magnetic field below the minimum energy value ($B/B_{\rm eq}=0.25$). The lobe pressure (Fig.~\ref{fig:prhot}) is $\simeq 10^{-10}\,{\rm dyn\,cm^{-2}}$ at $t\simeq2.5\,{\rm Myr}$, when the tip head is at $L_{\rm BS} \simeq 100\,{\rm kpc}$, and $\simeq 6\times10^{-11}\,{\rm dyn\,cm^{-2}}$ at $t\simeq5\,{\rm Myr}$, when $L_{\rm BS} \simeq 200\,{\rm kpc}$. 
  
  We observe that the lobe pressure in our simulation is of the same order than, or below, those given by \cite{od09} from estimates obtained for magnetic fields below equipartition, but typically larger than those given in \cite{ine17}. In the latter work, the authors also give the estimated lobe volumes (Figure 9 in that paper), which can be compared with the values shown in Fig.~\ref{fig:kine}. For sources with linear sizes around 100~kpc, the lobe volume is estimated to be around $10^5\,{\rm kpc^3}$, an order of magnitude over the value obtained from our simulation. The larger ICM pressure used in our simulation  at these scales (possibly caused by the group density profile considered), is certainly related to this difference, and can partially explain the discrepancy in lobe pressures between the simulation and the FRII jets in \cite{ine17}. A further effect that has to be taken into account is the power of these sources: Table~7 of \cite{ine17} shows the estimated jet powers for the studied objects, and these are between $10^{44}$ and $10^{45}\,{\rm erg/s}$, i.e., smaller than the power of J0. This means that the lobe pressure, which is proportional to the jet power and inversely proportional to the lobe volume, is necessarily smaller in the FRII jets considered. On the contrary, the sources studied by \cite{od09} are of the same power or larger than J0 (see Fig.~8 in that paper), and the authors report typically larger lobe pressures than our simulation. 
    
  It is relevant to state that, although the simulated cocoon volume is smaller than the typical values given for FRII sources \citep{ine17}, a linear fit $\log V_{\rm c}  \,-\,\log L_{\rm BS}$ for Phase II ($L_{\rm BS} \geq 80\,{\rm kpc}$) gives a slope of $2.75\pm0.08$, very close to that given by the fit to the observed values in Fig.~9 of \cite{ine17} ($2.61 \pm0.03$). The similarity between both slopes could be interpreted as the observed large-scale FRIIs being at the transition between the initial phase and the full three-dimensional phase, when the dentist-drill mechanism starts to play a relevant role. Actually, some of the sources show irregularities close to the jet head, which can be associated to helical oscillations \citep[see, e.g., 3C 223 and 3C 452, in][for an example]{har16,har17}. On the contrary, Phase I in the simulation gives a smaller slope, $2.07\pm0.04$. 

 A recent paper by \cite{sni18} gives estimates of ICM density and pressure for the FRII Cygnus~A which are an order of magnitude above those used in this work, and the jets have a power $L_{\rm k} = 6 - 8\times10^{45}\,{\rm erg/s}$, i.e., a factor between 6 and 8 larger than J0. Therefore, a higher lobe pressure in the source than in J0 could be expected. Although the authors only report on the post-shock pressures, the fast homogeneization of the pressure within the shocked region allows us to take this value as a close upper limit to the mean pressure, so $P_{\rm c,Cyg~A} \simeq 6 - 10 \times10^{-10} \,{\rm dyn\,cm^{-2}}$ at $L_{\rm BS}\leq 100\,{\rm kpc}$, which is indeed larger than the J0 value by close to an order of magnitude. Furthermore, the morphology of Cygnus~A is different from that of the FRII radio jets mentioned above, with Cygnus~A showing wider lobes. This is most probably due to the larger ambient density in which Cygnus~A jets evolve, which has probably had as a consequence the trigger of the 3D phase at smaller distances to the AGN, as shown by the large amplitude kink observed in the jets and the complex hot-spot structure \citep[clearly associated to the dentist drill effect][]{pyr15}.  
 
 \subsubsection{Hot-spot velocity and Mach number}
 
   \cite{od09} report on values of advance velocity for all the studied sources, which show no correlation with the linear size, as derived from spectral ages. The values of the expansion speed depend on the magnetic field considered (via the resulting spectral age). As in the case of lobe pressure, our results compare better to the lobe expansion speed obtained for a magnetic field below equipartition, $v_{\rm hs} = 0.01 - 0.1\,c$ (the velocities obtained for a magnetic field in equipartition with internal energy are all clearly above $0.1\,c$). On the contrary, \cite{har16} estimate typically smaller expansion speeds $v_{\rm hs} \simeq 0.01\,c$ for 3C~223 and 3C~452, also using spectral ages. In this case, we already noted that these jets might have developed already large scale helical oscillations that can decelerate the jet expansion. Nevertheless, the radio morphologies show thin cocoons with large aspect ratios, which is a consequence of fast axial expansion velocities: the smaller the ratio $v_{\rm j}/v_{\rm hs}$, the thinner the cocoon.  
 
  The simulated ICM has a sound speed of $2.3\times10^{-3}\,c$ (corresponding to $kT\simeq1.5\,{\rm keV}$ beyond $7.8\,{\rm kpc}$), which, combined with $v_{\rm hs}$, results in $M\simeq 50$ for the second phase. In contrast, \cite{har16} give small values, $M=1.8$, from the estimates of internal (lobe) and external pressure, and \cite{ine17} report $M \leq 5$ as lower limits of the Mach number for the FRII sources studied. \cite{ine17} also compute the shock Mach number by comparing the internal lobe pressure ($P_{\rm l}$ in Table~\ref{tab:tab3}) with the ICM pressure at the lobe tip ($P_{\rm ICM}$  in Table~\ref{tab:tab3}), via the expression \citep{wb06}:
\begin{equation} \label{eq:mach}
M^2\,=\,\frac{1}{2 \Gamma}\left( (\Gamma + 1) \frac{P_{\rm l}}{P_{\rm ICM}} + (\Gamma -1) \right).
\end{equation}  
If we use this equation to compute the shock Mach number, we obtain $M = 6.3$, which is still higher, but of the order of the values given in \cite{ine17} and reported in Table~\ref{tab:tab3}. Thus, the discrepancy is alleviated when taking into account that the shock Mach number is maximum at the jet head. In the case of Cygnus~A, where \cite{sni18} give $M\simeq 2.5$, the larger ICM temperature \citep[$kT \simeq 5\,{\rm keV}$, see Fig.~4 in][]{sni18} contributes to the reduction of the Mach number with respect to the simulation. 
    
 Summarizing, the FRII jets studied in \cite{od09}, \cite{har16} and \cite{ine17}, which we have used for comparison, are probably entering into the non-linear dominated expansion phase and thus being decelerated. However, their morphologies imply faster advance speeds in the previous phase, which we claim to be produced as a combination of the expansion through a dilute ambient medium and the small scale head oscillation (phase I). The amplitude of those oscillations grows via the coupling to large-scale instability modes, and this leads to the eventual large scale oscillation (dentist drill).

\subsection{Impact of the jet propagation on the ambient medium}

Despite the resolution limitations discussed in \ref{sec:res}, we consider that our 3D simulation is representative of a possible jet evolution through a hot galactic atmosphere. The fact that between 80~\% and 90~\% of the injected energy in relativistic jets is transferred to the ambient medium implies that a jet with power $10^{45}\,{\rm erg/s}$ injects $8-9\times 10^{44}\,{\rm erg/s}$ in the intracluster medium. This means that $2-3\times 10^{58}\,{\rm erg}$ ($\sim 10^{67}\,{\rm keV}$) are transferred every $10^6\,{\rm yr}$ to the ambient particles. If we consider a region of 100~kpc$^3$ and a source that has injected particles for 10~Myr, we find that a total of $\sim 10^{68}\,{\rm keV}$ are injected into $\sim 10^{70}\,{\rm cm^3}$. Taking into account that the intracluster number densities are $\ll 1 {\rm cm^{-3}}$, these numbers represent a large energy budget per particle (e.g., taking $n \sim 10^{-3}\,{\rm cm^{-3}}$, implying 10~keV per particle). Any reheating mechanism able to stop
the cooling flows would require energies per particle larger than $1\,{\rm keV}$ \citep{mn07}. 
Thus, the effects of the 3D jet would be a natural mechanism  to stop the cooling flows and provide the required energy.

 Furthermore, our result have important implications on the energetics of X-ray cavities as they are commonly computed: The work done by the jet to create the X-ray cavity is estimated as the product of the ambient pressure, obtained from X-ray data, times the volume of the X-ray cavity \citep[e.g.,][]{mc05}. The jet power is then obtained by means of dividing this work by the age of the radio-source, which is estimated assuming that the current size has been reached by buoyant motion ab initio. If the radio source has gone through a rapid expansion phase like the one captured by our simulations (and possibly taking place in powerful FRIIs, see the previous section), the age of the radio source can be severely overestimated, thus giving an underestimate of the jet power required to create the observed cavities. 

Radio relics are observed at large scales \citep[e.g.,][]{ens98,vw16,jh17}, implying a large population of energetic electrons within the intracluster medium. Our results show that, under some circumstances, relativistic jets can deposit a large amount of energetic particles to hundreds of kiloparsecs from the central source in a few million years. These particles eventually diffuse within the host clusters when the AGN is exhausted and thus provide a possible source for the energetic electrons observed in radio relics. Taking into account that radio relics are observed at low radio-frequencies, we can estimate the cooling times of the electrons injected in the intracluster medium by the jet using a frequency of, e.g., 178~MHz. Because our simulation is purely hydrodynamical, an equipartition magnetic field represents an upper limit of the field. In cgs units, the cocoon magnetic field $B_{\rm c} = \sqrt{8 \pi P_{\rm c}} \sim 0.01\,{\rm mG}$ for the last part of the simulation. With this field we estimate a cooling time of $\sim10^8\,{\rm yr}$, which would be even longer for magnetic fields below equipartition. Furthermore, we have to point out that the mean cocoon field drops with time due to expansion, thus keeping the cooling times long.

\subsection{Magnetic fields}

In this paper we have considered the evolution of purely hydrodynamical jets whose flux of energy are in the form of internal and kinetic energy fluxes. If the jets transport non-negligible magnetic fields, part of the total energy flux will be in the form of magnetic flux. At the hot-spot part of the magnetic energy could be transformed into internal energy through reconnection and then redistribute in the cocoon, thus contributing to the total cocoon's pressure. At the same time, if the jet is not magnetically dominated, the magnetic field surviving in the post-shock region will be advected down the cocoon backflow. Only if the magnetic field is dynamically dominant could it affect the characteristic instability growth scales within the jet hence modifying the dynamics of the jet head and its advance speed. Moreover, the sideways expansion of the cocoon would have to adapt to the action of the magnetic tension.

   Recent estimates based on observations \citep{ine17} show that jets are possibly particle dominated but still close to equipartition between the internal energy of the emitting particles and magnetic fields at kpc-scales. However these estimates do not take into account the contribution of the thermal component to the total particle energy in the jet because of the obvious difficulty to estimate it. Nevertheless, lobe pressure estimates \citep{cro04,cro05} require the presence of such a thermal component within the lobe gas in order to explain the lobe overpressure driving the observed shocks that surround AGN jets. Despite these considerations, future RMHD simulations should be run in order to test the effects of close-to-equipartition magnetic fields on the long-term evolution of jets and the jet/ambient medium feedback mechanism.
 
\section{Summary}

 In this paper we have focused on the long-term evolution in three dimensions of an FRII-type jet (kinetic luminosity $10^{45}$erg s$^{-1}$). Computational constraints limited us to follow the evolution until the jet propagated along 200~kpc (5.4~Myrs) with a resolution of 1 cell$/R_{\rm j}$ at injection. We have shown this resolution to be enough to properly describe i) the propagation of the jet through a density decreasing atmosphere in the case of axisymmetric simulations, and ii) the development (at their low-limit growth rates) of the three dimensional effects that determine the early evolution of the jet. The simulation presented here represents the largest 3D numerical simulation of a jet so far in terms of grid physical size and number of computational cells.

The simulation is part of a research plan aimed at characterizing the long term evolution of AGN jets and their cavities under increasingly realistic conditions. In this paper we have concentrated on the comparison with previous 2D axisymmetric simulations and simple analytical models to i) identify genuinely 3D effects affecting the long term evolution of jets, and ii) probe the efficiency of ambient heating by relativistic jets in 3D models. Three-dimensional effects in the jet propagation were triggered by adding a helical perturbation at the jet base.

The main results achieved in this paper are:

\begin{itemize}

\item The evolution of the jet proceeds along two well separated phases. In the initial phase the jet propagation is dominated by small 3D effects within the beam that increase the propagation efficiency momentarily. In the second phase, the interaction of the jet flow with internal (e.g., the injected perturbations) and/or cocoon-driven instabilities at the jet/cocoon contact discontinuity causes the deceleration of the jet advance. The comparison with both 2D axisymmetric simulations and simple analytical models support the validity of this description. 

\item The characteristics and duration of the initial phase depend strongly on the initial amplitudes and growth rates of the injected perturbations.

\item The 3D simulation confirms previous results based on 2D axisymmetric models on the efficiency of the shock-driven mechanism for the heating of the ambient medium.

\item Our simulation gives values for the lobe pressures and volumes, as well as the jet advance speed and shock Mach number, which match well with the properties of a sample of powerful radio jets of comparable size in similar environments. This result gives us confidence in our numerical approach towards a full understanding of the morphology, dynamics and feedback of powerful jets.

\end{itemize}

The present work reinforces the idea that a relativistic description of galactic jets is compulsory if a realistic portrait of the AGN feedback is desired. Our ongoing project includes the prolongation of the present simulation, together with the implementation of more realistic environments. Cooling processes, which could affect the stability and the strength of the bow shock, will be also a key ingredient of the new set of simulations.

\section*{Acknowledgements}
We thank the anonymous referee for his/her constructive criticism to the original manuscript. This work has been supported by the Spanish Ministerio de Econom\'{\i}a y Competitividad (grants AYA2015-66899-C2-1-P and AYA2016-77237-C3-3-P) and the Generalitat Valenciana (grant PROMETEOII/2014/069). Computer simulations have been carried out in the Red Espa\~nola de Supercomputaci\'on (Mare Nostrum and Tirant supercomputers) and in the Servei d'Inform\`atica de la Universitat de Val\`encia. The authors thank Carmen Aloy for the development of the parallelisation of the current version of the code.




\appendix

\subsection*{Appendix~A. Numerical convergence and reliability of simulation J0}

 The set up of our numerical simulations were described at the end of Sect.~2.2. As explained there, the numerical resolution of 1 cell per jet radius\footnote{Before proceeding, let us note that due to the ambient-density gradient, the jet in simulation J0 rapidly expands close to injection, reaching an effective resolution of 3 cells/$R_{\rm j}$ beyond 10~kpc.} of simulation J0 was chosen to compare with previous 2D axisymmetric simulations with the same numerical resolution, and also for computational constraints. Numerical simulations of jets usually resolve the jet inlet into 10 to 20 cells, but even in those cases the convergence is not perfect since the jet dynamics depends strongly on the interaction of the jet itself with the turbulent cocoon which is very sensitive to numerical diffusion \citep[e.g.,][]{mb05}. Besides that, most of the convergence studies \citep[e.g.,][]{mi10} are based on short-term simulations of jets propagating through homogeneous atmospheres. However, at this point, it is interesting to note that the convergence issue is less dramatic for the study of the long-term evolution of global quantities (our objective in this paper) since this study is based on large-scale structures with a large effective numerical resolution.

Besides the long-term simulation J0, we have performed two additional sets of simulations with increased resolutions of 2, 4 and 8  cells/$R_{\rm j}$. The first set corresponds to axisymmetric, 2D simulations propagating down the pressure and density decreasing ambient medium. The positions of the tip of the bow shock as pushed by the head of the jet during the first 2 Myrs of evolution (the duration of the so-called Phase I of the 3D simulation) are shown in the top panel of Fig.~\ref{fig:inner}. The simulations (including the original J45l one with 1 cell/$R_{\rm j}$), steered by the ambient density gradient, do show a clear convergence. In the same panel, the head propagation of the 3D jet simulation J0 departs from the converged axisymmetric solution due to the three-dimensional effects (the {\it shock-wobbling} described at Sect.~3.2) developing from the helical perturbation triggered at the jet base. A proof of this is shown in the middle panel of Fig.~\ref{fig:inner}. 

The second set of simulations are three-dimensional, perturbed simulations identical to J0 but with 2, 4 and 8  cells/$R_{\rm j}$ (J2, J4 and J8 in the paper, respectively). The positions of the head of the jet along the time for the first 0.4 Myrs corresponding to these simlations are shown at the bottom panel of Fig. ~\ref{fig:inner}. Convergence in the jet head position along the time is reached beyond 4 cells/$R_{\rm j}$ (see diamonds and triangles representing the head positions for J4 and J8 simulations in this plot). However, it is interesting to note that the effects that characterize the three-dimensional evolution tend to increase with increased resolution. As discussed in Sect.~3.2, the increase in the resolution improves the numerical representation of the terminal shock and its motions (i. e., the shock wobbling). Besides that, improved resolution allows for a faster growth of the instability amplitude which contributes to enhance the effect during the linear phase. As a result, the propagation efficiency of the jet in model J0 (with 1 cell/$R_{\rm j}$ at injection) represents a lower bound of the propagation efficiencies obtained at larger, converged resolutions (simulations J4 and J8).

In summary, the propagation of the jet through a density decreasing atmosphere reduces the demands of numerical resolution in the convergence under grid refinement which in the case of 2D axisymmetric simulations is reached even for 1 cell/$R_{\rm j}$. On the other hand, perturbed, 3D simulations need at least 4 cells/$R_{\rm j}$ for convergence. However, interestingly, the propagation efficiency of the jet prompted by the three-dimensional dynamics of the jet's terminal shock tends to increase with resolution leaving simulation J0 as a limiting case where to study a lower bound of the effect.

Finally, as pointed at the end of Sect.~3.2, the fact that the dynamics of the jet's terminal shock depend strongly on the initial amplitudes and growth rates of the perturbations triggered at the jet base means that there is not a unique evolution in three dimensions. However, the broad properties of this evolution (initial phase of enhanced jet propagation triggered by the propagation through a density-decreasing atmosphere and the 3D effects affecting the dynamics of the jet's terminal shock, and a long-term multidimensional phase), as shown in simulation J0, appear as robust.


\bsp	
\label{lastpage}
\end{document}